\documentclass[11pt]{article}

\usepackage{calc}
\usepackage{rotating}
\usepackage[english]{babel}
\usepackage[T1]{fontenc}
\usepackage[utf8]{inputenc}
\usepackage{graphicx}
\usepackage{subfig}
\usepackage{float}
\usepackage{amsmath}
\usepackage{amssymb}
\usepackage{amsthm}
\usepackage{latexsym}
\usepackage{dcolumn}
\usepackage{geometry}
\usepackage{color}
\usepackage{hyperref}

\newcommand{\newc}{\newcommand*}

\long\def\begincomment#1\endcomment{%
        \begingroup\sf\baselineskip12pt#1\endgroup}

\newc{\etal}{\textrm{et al.}} 
\newc{\eg}{\textrm{e.g.}} 
\newc{\ie}{\textrm{i.e.}}
\newc{\etc}{\textrm{etc}}
\newc\vs{\textrm{vs.}}
\newc{\cl}{\rm {CL}}

\newc{\ev}{\ensuremath{\,\mathrm{eV}}}
\newc{\kev}{\ensuremath{\,\mathrm{keV}}}
\newc{\mev}{\ensuremath{\,\mathrm{MeV}}}
\newc{\gev}{\ensuremath{\,\mathrm{GeV}}}
\newc{\tev}{\ensuremath{\,\mathrm{TeV}}}

\newc{\MeV}{\mev} 
\newc{\TeV}{\tev}
\newc{\invpb}{\ensuremath{/\text{pb}}}
\newc{\invfb}{\ensuremath{/\text{fb}}}

\newc\nb{\ensuremath{\,\mathrm{nb}}} \newc\pb{\ensuremath{\,\mathrm{pb}}} \newc\fb{\ensuremath{\,\mathrm{fb}}}

\newc\pc{\ensuremath{\,\mathrm{pc}}}
\newc\kpc{\ensuremath{\,\mathrm{kpc}}}
\newc\mpc{\ensuremath{\,\mathrm{Mpc}}}

\newc\ps{\ensuremath{\,\mathrm{ps}}} 

\newc\cmeter{\ensuremath{\,\mathrm{cm}}} 
\newc\meter{\ensuremath{\,\mathrm{m}}} 
\newc\kmeter{\ensuremath{\,\mathrm{km}}}

\newc\second{\ensuremath{\,\mathrm{s}}}
\newc\msecond{\ensuremath{\,\mathrm{ms}}}
\newc\nsecond{\ensuremath{\,\mathrm{ns}}}
\newc\psecond{\ensuremath{\,\mathrm{ps}}}

\newc{\chisqmin}{\ensuremath{\chi^2_{\mathrm{min}}}}
\newc{\Delchisq}{\ensuremath{\Delta\chi^2}}
\newc{\chisq}{\ensuremath{\chi^2}}

\newc{\like}{\ensuremath{\mathcal{L}}}

\newc\lsim{\ensuremath{\mathrel{\rlap{\lower4pt\hbox{\hskip1pt$\sim$}}\raise1pt\hbox{$<$}}}}
\newc\gsim{\ensuremath{\mathrel{\rlap{\lower4pt\hbox{\hskip1pt$\sim$}}\raise1pt\hbox{$>$}}}}

\newc{\VEV}[1]{\ensuremath{\langle #1 \rangle}}

\newc{\dl}{\ensuremath{\stackrel{\leftarrow}{D}}}
\newc{\dr}{\ensuremath{\stackrel{\rightarrow}{D}}}

\newc{\bcenter}{\begin{center}}    \newc{\ecenter}{\end{center}}
\newc{\bfl}{\begin{flushleft}}    \newc{\efl}{\end{flushleft}}
\newc{\bfr}{\begin{flushright}}    \newc{\efr}{\end{flushright}}

\newc{\bi}{\begin{itemize}}
\newc{\ei}{\end{itemize}}
\newc{\bed}{\begin{description}}
\newc{\eed}{\end{description}}
\newc{\ben}{\begin{enumerate}}
\newc{\een}{\end{enumerate}}

\newc{\be}{\begin{equation}}
\newc{\ee}{\end{equation}}
\newc{\bea}{\begin{eqnarray}}
\newc{\eea}{\end{eqnarray}}
\newc{\ra}{\rightarrow}

\newc{\alphas}{\ensuremath{\alpha_s}}
\newc{\alphatwo}{\ensuremath{\alpha_2}}
\newc{\alphaone}{\ensuremath{\alpha_1}}
\newc{\alphai}[1]{\ensuremath{\alpha_{#1}}}
\newc{\alphaem}{\ensuremath{\alpha_{\mathrm{em}}}}

\newc{\alphaeff}{\ensuremath{\alpha_{\mathrm{eff}}}}

\newc{\sineff}{\ensuremath{\sin \theta_{\mathrm{eff}}}}
\newc{\sinsqeff}{\ensuremath{\sin^2 \theta_{\mathrm{eff}}}}
\newc{\dalphahad}{\ensuremath{\Delta \alpha_{\mathrm{had}}}}

\newc{\yt}{\ensuremath{h_t}} \newc{\yb}{\ensuremath{h_b}} \newc{\ytau}{\ensuremath{h_{\tau}}}

\newc\mz{\ensuremath{M_Z}} 
\newc\mw{\ensuremath{m_W}}
\newc\mZ{\mz}        \newc\mW{\mw}

\newc\mhsm{\ensuremath{ m_{H_{\mathrm{SM}}}}}

\newc{\mtop}{\ensuremath{ m_t}}               \newc{\mtpole}{\ensuremath{ M_t}}
\newc{\mbottom}{\ensuremath{ m_b}} 
\newc{\mtau}{\ensuremath{ m_{\tau}}}
\newc{\mt}{\mtpole}
\newc{\mb}{\mbottom} 

\newc{\honesm}{\ensuremath{H_1=H_{\rm SM}}}
\newc{\htwosm}{\ensuremath{H_2=H_{\rm SM}}}

\newc{\muZZ}{\ensuremath{\mu_{ZZ}}}
\newc{\mugaga}{\ensuremath{\mu_{\gamma\gamma}}}

\newc{\hsm}{\ensuremath{H_{\rm SM}}}

\newc{\rtwogg}{\ensuremath{R_{h_2}(\gamma\gamma)}}
\newc{\rtwozz}{\ensuremath{R_{h_2}(ZZ)}}
\newc{\ronegg}{\ensuremath{R_{h_1}(\gamma\gamma)}}
\newc{\ronezz}{\ensuremath{R_{h_1}(ZZ)}}
\newc{\rsiggg}{\ensuremath{R_{h_\textrm{sig}}(\gamma\gamma)}}
\newc{\rsigzz}{\ensuremath{R_{h_\textrm{sig}}(ZZ)}}

\newc{\llbar}{\ensuremath{\ell\bar{\ell}}}
\newc{\tauptaum}{\ensuremath{ \tau^+\tau^-}}

\newc{\qqbar}{\ensuremath{ q\bar{q}}} \newc{\ppbar}{\ensuremath{ p\bar{p}}}
\newc{\bbbar}{\ensuremath{ b\bar{b}}} \newc{\ttbar}{\ensuremath{ t\bar{t}}}
\newc{\ffbar}{\ensuremath{ f\bar{f}}} \newc{\tautaubar}{\ensuremath{ \tau\bar{\tau}}}

\newc{\mchi}{\ensuremath{m_\neutone}}
\newc{\squark}{\ensuremath{\tilde{q}}}
\newc{\slepton}{\ensuremath{\tilde{l}}}
\newc{\gluino}{\ensuremath{\tilde{g}}} 
\newc{\mgluino}{\ensuremath{{m_{\gluino}}}}

\newc{\sthw}{\ensuremath{ \sin\theta_W}}              \newc{\cthw}{\ensuremath{\cos\theta_W}}
\newc{\tanthw}{\ensuremath{ \tan\theta_W}}              \newc{\cotthw}{\ensuremath{\cot\theta_W}}

\newc{\ssqthw}{\ensuremath{\sin^2 \theta_W}}

\newc{\msbar}{\ensuremath{\overline{MS}}} \newc{\drbar}{\ensuremath{\overline{DR}}}

\newc{\mtmtsmmsbar}{\ensuremath{ m_t(m_t)^{\msbar}_{{\mathrm{SM}}}}}
\newc{\mtmtsmdrbar}{\ensuremath{ m_t(m_t)^{\drbar}_{{\mathrm{SM}}}}}
\newc{\mtmtmssmdrbar}{\ensuremath{ m_t(m_t)^{\drbar}_{{\mathrm{SUSY}}}}}

\newc{\mbmbmsbar}{\ensuremath{ m_b(m_b)^{\msbar} }}

\newc{\mbmbsmmsbar}{\ensuremath{ m_b(m_b)^{\msbar}_{{\mathrm{SM}}}}}
\newc{\mbmzsmmsbar}{\ensuremath{ m_b(\mz)^{\msbar}_{{\mathrm{SM}}}}}
\newc{\mbmzsmdrbar}{\ensuremath{ m_b(\mz)^{\drbar}_{{\mathrm{SM}}}}}
\newc{\mbmzmssmdrbar}{\ensuremath{ m_b(\mz)^{\drbar}_{{\mathrm{SUSY}}}}}

\newc{\mtaumzsmmsbar}{\ensuremath{ m_{\tau}(\mz)^{\msbar}_{{\mathrm{SM}}}}}
\newc{\mtaumzsmdrbar}{\ensuremath{ m_{\tau}(\mz)^{\drbar}_{{\mathrm{SM}}}}}
\newc{\mtaumzmssmdrbar}{\ensuremath{ m_{\tau}(\mz)^{\drbar}_{{\mathrm{SUSY}}}}}

\newc{\alphasmzms}{\ensuremath{\alpha_s(M_Z)^{\overline{MS}}}}
\newc{\alphaimzms}[1]{\ensuremath{\alpha_{#1}(M_Z)^{\overline{MS}}}}

\newc{\alphaemmz}{\ensuremath{\alpha_{\mathrm{em}}(M_Z)^{\overline{MS}}}}

\newc{\mzero}{\ensuremath{{m_0}}}
\newc{\mhalf}{\ensuremath{ m_{1/2}}}
\newc{\tanb}{\ensuremath{\tan\beta}}
\newc{\azero}{\ensuremath{ A_0}}
\newc{\signmu}{\ensuremath{\rm{sgn}\,\mu}}
\newc{\atau}{\ensuremath{{A_{\tau}}}}

\newc{\mueff}{\ensuremath{\mu_{\rm{eff}}}}

\newc{\lam}{\ensuremath{{\lambda}}}
\newc{\kap}{\ensuremath{{\kappa}}}

\newc{\alam}{\ensuremath{{A_{\lambda}}}}
\newc{\akap}{\ensuremath{{A_{\kappa}}}}

 \newc{\hs}{\ensuremath{ H_s}}      
\newc{\mhs}{\ensuremath{ m_{H_s}}} 

\newc{\mgut}{\ensuremath{ M_{\rm GUT}}}
\newc{\mplanck}{\ensuremath{ M_{\rm P}}}      \newc{\mpl}{\ensuremath{ M_{\rm Pl}}}
\newc{\msusy}{\ensuremath{ M_{\rm SUSY}}}      \newc{\ms}{\ensuremath{ M_{\rm S}}}

 \newc{\hu}{\ensuremath{ H_u}}       \newc{\hd}{\ensuremath{ H_d}}
 \newc{\mhu}{\ensuremath{ m_{H_u}}}       \newc{\mhd}{\ensuremath{ m_{H_d}}}
 \newc{\mhuew}{\ensuremath{ m^{\ast}_{H_u}}}       \newc{\mhdew}{\ensuremath{ m^{\ast}_{H_d}}}
 \newc{\mhuewsq}{\ensuremath{ m^{\ast\, 2}_{H_u}}}       \newc{\mhdewsq}{\ensuremath{ m^{\ast\, 2}_{H_d}}}
 \newc{\mhl}{\ensuremath{m_\hl}} 
 \newc{\mhone}{\ensuremath{m_{H_1}}} 
 \newc{\mhtwo}{\ensuremath{m_{H_2}}} 
 \newc{\mglu}{\ensuremath{m_{\tilde g}}} 
 \newc{\mul}{\ensuremath{m_{\tilde{u}_L}}} 
 \newc{\mtone}{\ensuremath{m_{\tilde{t}_1}}} 
 \newc{\ma}{\ensuremath{m_A}} 
 \newc{\maone}{\ensuremath{m_{A_1}}} 
 \newc{\matwo}{\ensuremath{m_{A_2}}}
 \newc{\hone}{\ensuremath{H_1}}
 \newc{\htwo}{\ensuremath{H_2}}
 \newc{\aone}{\ensuremath{A_1}}
 \newc{\atwo}{\ensuremath{A_2}}

\newc{\sigsip}{\ensuremath{\sigma^{\rm SI}_{p}}}	\newc{\sigsin}{\ensuremath{\sigma^{\rm SI}_{n}}}
\newc{\sigsdp}{\ensuremath{\sigma^{\rm SD}_{p}}}	\newc{\sigsdn}{\ensuremath{\sigma^{\rm SD}_{n}}}
\newc{\sigsi}{\ensuremath{\sigma^{\rm SI}}}	\newc{\sigsd}{\ensuremath{\sigma^{\rm SD}}}

\newc{\abund}{\ensuremath{ \Omega h^2}}
\newc{\omegadm}{\ensuremath{ \Omega_{{\rm DM}}}}     \newc{\abunddm}{\ensuremath{ \Omega_{{\rm DM}} h^2}} 
\newc{\omegam}{\ensuremath{ \Omega_{{\rm m}}}}       \newc{\abundm}{\ensuremath{ \Omega_{{\rm m}} h^2}}
\newc{\omegab}{\ensuremath{ \Omega_{{\rm b}}}}	\newc{\abundb}{\ensuremath{ \Omega_{{\rm b}} h^2}}
\newc{\omegatot}{\ensuremath{ \Omega_{{\rm TOT}}}}
\newc{\omegacdm}{\ensuremath{ \Omega_{{\rm CDM}}}}   \newc{\abundcdm}{\ensuremath{ \Omega_{{\rm CDM}} h^2}}
\newc{\omegalambda}{\ensuremath{ \Omega_{\Lambda}}} \newc{\abundlambda}{\ensuremath{ \Omega_{\Lambda} h^2}}
\newc{\omegarad}{\ensuremath{ \Omega_{{\rm rad}}}}  \newc{\abundrad}{\ensuremath{ \Omega_{{\rm rad}} h^2}}

\newc{\rhocrit}{\ensuremath{ \rho_{\rm crit}}}
\newc{\rhochi}{\ensuremath{ \rho_{\chi}}}

\newc{\abunchi}{\ensuremath{\Omega_\chi h^2}}
\newc{\abundlsp}{\ensuremath{\Omega_{\rm LSP}h^2}}

\newc{\amu}{\ensuremath{ a_{\mu}}}        \newc{\amususy}{\ensuremath{ a_{\mu}^{\mathrm{SUSY}}}}
\newc{\amuexpt}{\ensuremath{ a_{\mu}^{\mathrm{expt}}}}        \newc{\amusm}{\ensuremath{ a_{\mu}^{\mathrm{SM}}}}
\newc\deltaamu{\ensuremath{\Delta a_{\mu}}} \newc{\deltaamususy}{\ensuremath{\delta a_{\mu}^{\mathrm{SUSY}}}}
\newc\gmtwo{\ensuremath{ (g-2)_{\mu}}} 
\newc{\deltagmtwomususy}{\ensuremath{\delta\left(g-2\right)_{\mu}^{\mathrm{SUSY}}}}
\newc{\deltagmtwomu}{\ensuremath{\delta\left(g-2\right)_{\mu}}}

\newc\BR{\ensuremath{\rm BR}}

\newc\bsgamma{\ensuremath{ b\rightarrow s \gamma }}
\newc\bxsgamma{\ensuremath{\overline{B}\rightarrow X_{s}\gamma}}

\newc\brbsgamma{\ensuremath{\BR\left(\bsgamma\right)}}
\newc\brbxsgamma{\ensuremath{\BR\left(\bxsgamma\right)}}

\newc\bsmumu{\ensuremath{B_s\to\mu^+\mu^-}}
\newc\brbsmumu{\ensuremath{\BR\left(B_s\to\mu^+\mu^-\right)}}

\newc\bdmmumu{\ensuremath{\overline{B}_d\to\mu^+\mu^-}}

\newc\bbbarmix{\ensuremath{\overline{B}_s\mbox{-}B_s}}      
\newc\delmbs{\ensuremath{\Delta M_{B_s}}}

\newc{\butaunu}{\ensuremath{B_u \rightarrow \tau \nu}}
\newc{\brbutaunu}{\ensuremath{\BR\left(B_u \rightarrow \tau \nu\right)}}

\newcommand*{\neutone}{\ensuremath{\chi}}

\newcommand*{\micromegas}{MicrOMEGAs}

\newcommand*{\superiso}{\text{SuperIso}}

\newcommand*{\nmssmtools}{\text{NMSSMTools}}

\let\oldcite\cite
\renewcommand*{\cite}{~\oldcite}

\newcommand*{\hl}{\ensuremath{h}}

\begin{document}
\renewcommand{\thefootnote}{\fnsymbol{footnote}}
\begin{center}
{\large {\bf Detection prospects of light NMSSM Higgs pseudoscalar via
  cascades of heavier scalars from vector boson fusion and
  Higgs-strahlung}}\\
\vspace*{1cm}
{\bf N.-E.\ Bomark$^{a,b}$\footnote{nilserik.bomark@gmail.com}, S.\
  Moretti$^c$\footnote{S.Moretti@soton.ac.uk}, L.\
  Roszkowski$^{b,d}$\footnote{Leszek.Roszkowski@fuw.edu.pl}} \\
\vspace{0.3cm}
${}^a$Department of Natural Sciences, University of Agder, Postboks 422, 4604 Kristiansand, Norway\\
${}^b$National Centre for Nuclear Research, Ho\.za 69, 00-681 Warsaw, Poland\\
${}^c$School of Physics \& Astronomy, University of Southampton, Southampton SO17 1BJ, UK\\
${}^d$School of Physics \& Astronomy, University of Sheffield, S3 7RH, UK
\end{center}
\renewcommand{\thefootnote}{\arabic{footnote}}
\setcounter{footnote}{0}
%
%
\begin{abstract}
  A detection at the Large Hadron Collider of a light Higgs
  pseudoscalar would, if interpreted in a supersymmetric framework, be a smoking gun signature of non-minimal
  supersymmetry. In this work in the framework of the Next-to-Minimal
  Supersymmetric Standard Model we focus on vector boson fusion and
  Higgs-strahlung production of heavier scalars that subsequently
  decay into pairs of light pseudoscalars. We demonstrate that although these
  channels have in general very limited reach, they are viable for the detection of light pseudoscalars in some parts of parameter space and can serve as an important
  complementary probe to the dominant gluon-fusion production
  mode. We also demonstrate that in a Higgsfactory these channels may reach sensitivities comparable to or even exceeding the gluon fusion channels at the LHC, thus possibly rendering this our best option to discover a light pseudoscalar. It is also worth mentioning that for the singlet dominated scalar this may be the only way
  to measure its couplings to gauge bosons.
  Especially promising are channels where the initial scalar is
  radiated off a $W$ as these events have relatively high rates and
  provide substantial background suppression due to leptons from the
  $W$. We identify three benchmark points that well represent the
  above scenarios. Assuming that the masses of the scalars and
  pseudoscalars are already measured in the gluon-fusion channel, the event kinematics can be further constrained,
  hence significantly improving detection
  prospects. This is especially important in the Higgs-strahlung channels with rather
  heavy scalars, and results in possible detection at 200/fb for the most favoured
  parts of the parameter space.
\end{abstract}


\section{Introduction}
The presence of an extra singlet superfield in the Next-to-Minimal
Supersymmetric Standard Model (NMSSM) (see, e.g.,\cite{Ellwanger:2009dp} for a review) as compared to the MSSM, has a
significant impact on the ensuing phenomenology of the Higgs sector at
the Large Hadron Collider (LHC).  In particular, the NMSSM Higgs
sector is enlarged by two neutral mass eigenstates, one scalar and one
pseudoscalar, in addition to the three MSSM-like
ones.\footnote{Hereafter, our book-keeping of the physical Higgs
  states of the NMSSM will be as follows: the CP-even states will be
  denoted by $H_i$ (with $i=1,2,3$ and such that $ m_{H_1}< m_{H_2}<
  m_{H_3}$) while the CP-odd ones by $A_i$ (with $i=1,2$ and such that
  $ m_{A_1}< m_{A_2}$). We also use $\hsm$ for the scalar playing the role of the discovered Higgs boson and $H_S$ for the singlet dominated scalar.}  The singlet nature of the scalar component
of the additional superfield allows $H_1$ and $A_1$ to be very light when singlet dominated, even down to a few GeV, without
entering into conflict with current theoretical and experimental
constraints.  This is so because their couplings to the fermions and
gauge bosons of the SM are typically much smaller than those of the
doublet-dominated Higgs bosons ($H_{2,3}$ and $A_2$), which are
assumed heavier.  As a consequence, the observation of any of these
potentially light states, in addition to the SM-like Higgs boson
already discovered at the LHC\cite{Aad:2012tfa,Chatrchyan:2012ufa},
would constitute a hallmark signature of a non-minimal nature of
supersymmetry (SUSY). Careful examination of their mass and coupling
values in relation to the mass and coupling measurements of the ~125
GeV SM-like Higgs boson (and possibly other discovered Higgs states)
in a variety of production and decay channels will eventually enable
one to profile their possible NMSSM nature.

The lightest pseudoscalar $A_1$, in particular, can be the most crucial
probe of the NMSSM as it can be very light, so that in principle it is
accessible in meson decays, where it has been searched for initially\cite{Love:2008aa,Tung:2008gd,
  Lees:2012iw,Lees:2012te,Lees:2013vuj,Peruzzi:2014ksa}. The $A_1$
state with mass $\gtrsim 5$\gev\ has also been probed in the possible
decay of a heavy (SM-like or not) scalar Higgs boson at LEP2\cite{Abbiendi:2002in,Schael:2010aw}
and Tevatron\cite{Abazov:2009yi}, where no significant excess was
observed. Regarding the LHC, the situation is as follows.  CMS
searched for a light pseudoscalar produced either singly
\cite{Chatrchyan:2012am} or in pairs from the decays of a non-SM-like
Higgs boson\cite{Chatrchyan:2012cg} and decaying into the $\mu^+\mu^-$
channel, while the $A_1A_1\to4\tau$ signature (via a SM-like Higgs
decay) has been tackled in \cite{Khachatryan:2015nba}.\footnote{The sensitivity of the di-photon
  sample to a singly produced Higgs boson (of mass 150 GeV and above)
  decaying to $\gamma\gamma$ has also been investigated by CMS\cite{CMS:2014onr} alongside that of the $4b$ sample to pair
  production of ~125 GeV Higgs bosons\cite{CMS:2014eda}.}  For completeness, let us also mention an ATLAS search for scalar particles
decaying via narrow resonances into two photons in the mass range
above 65 GeV\cite{Aad:2014ioa}, though this does not constrain light pseudoscalars very much.

In addition, there are plenty of phenomenological analyses aiming at
assessing the scope of $A_1$ discovery within the NMSSM at the
LHC. Prior to the SM-like Higgs boson discovery, quite some effort was
put into extending the so-called `no-lose theorem' of the MSSM ---
stating that at least one Higgs boson of the MSSM would have been
discovered at the LHC via the usual SM-like production and decay
channels throughout the entire MSSM parameter space\cite{Dai:1993at}
(see\cite{Ellwanger:2013ova} for a recent review) to the case of
the NMSSM\cite{Accomando:2006ga,Almarashi:2010jm,Almarashi:2011bf,Ellwanger:2005uu,Ellwanger:2003jt,Ellwanger:2001iw,Hugonie:2001ib,Belyaev:2008gj,Forshaw:2007ra,Belyaev:2010ka,Moretti:2006sv,Moretti:2006hq,Almarashi:2011hj,Ellwanger:1999ji,Djouadi:2008uw,Mahmoudi:2010xp}.
In the light of the recent Higgs boson discovery though, the above
theorem is necessarily verified and, if one wants to prove the NMSSM
to be a viable alternative  to the MSSM, one ought to probe it away
from the MSSM limit.

Following this line of reasoning, if one abandons the limiting
case of SM-like decay channels of Higgs states, the NMSSM offers
interesting signatures which are precluded in the MSSM after the
latest experimental constraints, in the form of a variety of Higgs
$\to$ Two-Higgs and Higgs $\to$ Gauge-Higgs decays. A large volume of
phenomenological literature exists on these topologies, claiming that,
for certain NMSSM parameter choices, these would be accessible at the
LHC, eventually enabling one to disentangle the NMSSM from the MSSM
hypothesis, thereby establishing a so-to-say `more-to-gain'
theorem\cite{Moretti:2006sv}. The importance of such decays in the context of
the NMSSM has been emphasised in
Refs.\cite{Dermisek:2005ar,Gunion:1996fb,Dobrescu:2000jt} from the
point of view of fine-tuning as well as a distinctive NMSSM signature
at colliders. In particular, the $H_{1,2}\to A_1A_1$ mode has received
much attention. This decay can in fact be dominant in large regions of
the NMSSM parameter space.  It was realised that vector boson fusion
(VBF) could be a viable production channel to detect the above modes
at the LHC, in which the CP-odd Higgs pair decays into
$jj\tau^+\tau^-$\cite{Ellwanger:2005uu,Ellwanger:2003jt,Ellwanger:2004gz} (where $j$
represents a jet).  Some scope could also be afforded by a $4\tau$
signature in both VBF and Higgs-strahlung (HS) off-gauge bosons\cite{Belyaev:2008gj}.
The gluon-fusion (GF) channel too could be a
means of accessing $H_1\to A_1A_1$ decays, as long as the light CP-odd
Higgs states both decay into four muons\cite{Belyaev:2010ka} or two
muons and two taus\cite{Lisanti:2009uy}. Finally, the scope of NMSSM
neutral Higgs boson production in association with $b\bar b$ pairs was
assessed in\cite{Almarashi:2010jm,Almarashi:2011hj,Almarashi:2011bf,Almarashi:2011te,Almarashi:2011qq,Almarashi:2012ri},
including the case of $H_2\to ZA_1$ decays with $ Z\to jj$ and
$A_1\to\tau^+\tau^-$.

All the above mentioned analyses were, however, performed prior to the
discovery of the Higgs boson at the
LHC\cite{Chatrchyan:2012ufa,Aad:2012tfa}.  In the aftermath of the
discovery, detection prospects of all NMSSM Higgs bosons, including
also via their decays into other Higgs states, were recently
investigated in\cite{King:2014xwa}, though limited to the case of
inclusive rates. Further, in\cite{Kim:2012az}, the
$A_1\rightarrow\gamma\gamma$ decay channel was studied in the regime
of a light $A_1$. In \cite{Munir:2013wka} it was then noted that in
the NMSSM the $A_1$ could in fact be degenerate in mass with the
SM-like Higgs boson $H_{\rm SM}$. It could thus cause an enhancement in the Higgs boson
signal rates near 125\,\gev\ in the $\gamma\gamma$, $b\bar{b}$ and
$\tau^+\tau^-$ channels simultaneously, provided that it is produced
in association with a $b\bar{b}$ pair. The $b\bar b A_1$ channel was
also studied in detail in\cite{Kozaczuk:2015bea}, with a more
optimistic conclusion as compared to\cite{Bomark:2014gya}.  
In \cite{Cerdeno:2013cz} the $H_{\rm SM} \rightarrow A_1A_1
\rightarrow 4\ell$ (with $\ell$ denoting $e^\pm$ and $\mu^\pm$)
process at the LHC was studied in detail, while the $b\bar b\mu\mu$
final state was deemed the most promising in\cite{Curtin:2014pda}. The
production of $A_1$ via neutralino decays has also been recently
revisited in\cite{Stal:2011cz,Das:2012rr,Cerdeno:2013qta}.  Finally,
NMSSM benchmark proposals capturing much of this phenomenology exist
in the literature\cite{Djouadi:2008uw,King:2012tr}.

In this work we continue to pursue a recently started endeavour\cite{Bomark:2014gya},
with an intention to systematically analyse all
the production and decay processes that could potentially lead to the
detection of a light NMSSM $A_1$ at the LHC with $\sqrt{s} =14$\tev.
In\cite{Bomark:2014gya} (see also\cite{Bomark:2014qua,HPNP2015}), we
considered the case of the light pseudoscalar $A_1$ produced from a
heavy Higgs boson coming from GF which then decays into either $A_1$
pairs or $ZA_1$ (with the $Z$ in turn decaying into
electrons/muons). We found that the $A_1$ can be accessible through a
variety of signatures proceeding via $A_1\to \tau^+\tau^-$ and/or
$b\bar b$, the former assuming hadronic decays and the latter two
$b$-tags within a fat jet or two separate slim ones. Some of these
channels were also studied in the comprehensive review of exotic Higgs
decays contained in\cite{Curtin:2013fra}.

In the present paper, working under the assumption that a light $A_1$ state
has been found through one or more of the decay  channels analysed
in\cite{Bomark:2014gya}, we assess the scope of the LHC Run II in
profiling its nature by resorting instead to the VBF and HS Higgs
production modes. This  would
then enable access to the heavy Higgs couplings to both charged and
neutral gauge bosons, thus complementing the GF channel which only
allows one to measure their fermion couplings. It is worth emphasising that for the non-SM-like
scalars, this might be the only chance to access these couplings as the decay to pairs of
pseudoscalars can be completely dominating. Furthermore, although
the VBF and HS channels have significantly smaller cross sections than
GF, the improved possibilities for tagging through the additional
(forward/backward) jets for VBF and the additional leptons from vector
boson decays in HS, may still render them competitive against GF, for
which many of the triggers needed for fully hadronic final states have
been scarcely tested in the experimental environment. In view of this,
in this paper we perform a preliminary study of  light $A_1$s in 
the VBF and HS Higgs
production modes by making some simplifying assumptions, in particular
by ignoring detector effects and trigger thresholds.

Before plunging into the details of this new analysis, we should also
point out here that in the NMSSM both $H_1$ and $H_2$, the lightest
and next-to-lightest CP-even Higgs bosons, respectively, can
alternatively play the role of the SM-like Higgs boson $H_{\rm SM}$,
as emphasised already in\cite{Ellwanger:2011aa,King:2012is,Ellwanger:2012ke,Gherghetta:2012gb,Cao:2012fz}
and confirmed in\cite{Bomark:2014gya,Bomark:2014qua,HPNP2015}.

The article is organised as follows. In section~\ref{model}, we
define the parameter space of the NMSSM under
consideration.  In section~\ref{scans} we explain our approach to scan
the NMSSM parameter space while in section~\ref{analysis} we
describe our signal-to-background analyses. Then in
sections~\ref{VBF}, \ref{ZH} and \ref{WH} we discuss in
detail our results for VBF and HS (the latter separately for the neutral ($ZH$)
and charged ($WH$) channels) at the LHC over the entire NMSSM parameter space.  Then, after presenting some
benchmark points available for experimental investigation in section
 \ref{benchmarks} and testing them at the CERN $pp$ machine, we
afford a brief digression on the physics of a light $A_1$ state at a future $e^+e^-$ collider (in section \ref{epem}). We
 summarise and conclude in
section \ref{conclusions}.

\section{\label{model} The NMSSM parameter space}
The idea behind the NMSSM is to
explain the peculiar feature of the MSSM that the supersymmetry
preserving $\mu$ term is phenomenologically required to be of the
same scale as the soft supersymmetry breaking mass parameters, while
in principle it would be expected to be of a completely different
origin.

In the NMSSM this so-called $\mu$ problem is solved by introducing an extra gauge-singlet
chiral superfield $\widehat{S}$ whose scalar component receives a
vacuum expectation value (VEV) due to its soft supersymmetry breaking
terms. All that is needed to generate an effective $\mu$ term of the same order as the
soft supersymmetry-breaking parameters, is then to have a term
$\lambda\widehat{S}\widehat{H}_u\widehat{H}_d$ in the superpotential
and the (effective) $\mu$ term $\mueff\equiv \lam s$ will be given by the VEV of $S$
times the coupling constant \lam. We also need to add a cubic term in $\widehat{S}$ to the
superpotential, so that the terms involving $\widehat S$ read
\begin{equation}\label{eq:SuperPot}
  \lam \widehat{S}\widehat{H}_u\widehat{H}_d + \frac{\kap}{3}\widehat S^3,
\end{equation}
where \lam\ and \kap\ are dimensionless coupling constants. Further,
one needs to add the corresponding soft supersymmetry breaking
terms in the scalar potential. The soft supersymmetry breaking terms
relevant for the Higgs sector are:
\begin{equation}\label{eq:SoftHiggs}
  \mhu^2|H_u|^2+\mhd^2|H_d|^2+m_S^2|S|^2+(\lam \alam SH_uH_d + \frac{\kap}{3}\akap S^3+h.c.),
\end{equation}
where $\mhu$, $\mhd$, $m_S$, \alam\ and \akap\ are dimensionful mass
and trilinear parameters. By minimising the scalar potential we can
trade the three scalar mass parameters for \mueff\ and \tanb\
(i.e., the ratio between the VEVs in the up type and down type Higgs
doublets, $v_u/v_d$). We also replace \akap\ with $M_p$, the singlet pseudoscalar mass entry; if input is given at the Electroweak scale, this is close to the lightest pseudoscalar mass which improves the efficiency of the scan. For this reason we use parameters defined at the EW scale throughout the paper.

For the sfermion masses we use a common mass parameter \mzero\ and for all gaugino masses we use a common
parameter \mhalf\  but to take into account the effects of running from a high scale unification we use $M_2=\mhalf,\ M_1=\mhalf/2,\ M_3=3\mhalf$. Similarly, we use a common parameter \azero\ for all
trilinear parameters except \alam\ and \akap. Since the most significant effect these parameters
have on the Higgs sector is radiative corrections to the Higgs mass, unifying them in the
above manner should not impact much on the observables of interest, thus giving maximum freedom in the Higgs sector while
keeping the number of free parameters at a manageable level. This
leaves us with nine parameters:
\begin{equation}\label{eq:Params}
  \mzero,\ \mhalf,\ \azero,\ \tanb,\ \mueff,\ \lam,\ \kap,\ \alam,\ M_p.
\end{equation}

As can be seen in the tree-level mass for the lightest doublet scalar, i.e.\ essentially (up to mixing with the singlet scalar) $H_{\rm SM}$ in
the NMSSM\cite{Ellwanger:2009dp},
\begin{equation}
\label{eq:mhlim}
m_{H_{\rm SM}}^2 \simeq m_Z^2\cos^2 2\beta
 + \lam^2 v^2 \sin^2 2\beta \,,
\end{equation}
there is an additional contribution to the Higgs mass coming from the
\lam\ term, not present in the MSSM. This means that, when this term
is sizeable, i.e.,\ when \lam\ is large and \tanb\ is small, one can
obtain the measured ~125 GeV mass without resorting to large radiative
corrections from the stop sector, which is necessary in the MSSM. In the
forthcoming sections we will refer to this part of parameter space as
the `naturalness limit'. Also notice that, when \htwosm, it is possible that some
mixing between $H_1$ and $H_2$ increases \mhtwo\ further, hence making
it even easier to achieve (and indeed exceed) the required ~125
GeV\cite{Ellwanger:2011aa,Jeong:2012ma,Agashe:2012zq,Barbieri:2013hxa,Badziak:2013bda,Barbieri:2013nka}.

\section{\label{scans} The scans}
In order to investigate the prospects of the discussed channels in the NMSSM we performed a number of Bayesian scans of the above mentioned 9-dimensional parameter space. These scans use MultiNest-v2.18\cite{Feroz:2008xx} for nested sampling of the
parameter space and \nmssmtools -v4.6.0\cite{NMSSMTools}, including the bug-fix of version 4.7.1,  for the
calculation of the Higgs mass spectrum, couplings, Branching Ratios
(BRs) and constraints on the parameter points (including LEP-II bounds on light scalars and pseudoscalars). The output from
\nmssmtools\ is further processed by
HiggsBounds-v4.1.3\cite{Bechtle:2008jh,Bechtle:2011sb,Bechtle:2013gu,Bechtle:2013wla}
to ensure against exclusion from searches for other Higgs bosons. Also
\superiso -v3.3\cite{superiso} is used to calculate $b$-physics
variables. These are then required to comply with the constraints
from\cite{Beringer:1900zz} (in all cases, the last number refers to a theoretical uncertainty in the numerical evaluation):
\begin{itemize}
\item $\brbsmumu = (3.2\pm1.35\pm 0.32) \times 10^{-9}$,
\item $\brbutaunu = (1.66\pm 0.66 \pm 0.38) \times 10^{-4}$,
\item $\brbxsgamma = (3.43\pm 0.22 \pm0.21) \times
10^{-4}$.
\end{itemize}

To guard against over-closure of the Universe, an
upper bound of $\Omega_\chi h^2 < 0.131$ on the dark matter relic
density was also applied. This was set assuming a 10\% theoretical error on the
central value of 0.119 from PLANCK\cite{Ade:2013zuv} and the relic
density was calculated with the help of
\micromegas -v2.4.5\cite{micromegas}. We do not include any constraint on $(g-2)_\mu$; the contributions here are always too small to cause problems, they are actually too small to reconcile the experimental value with the Standard Model calculations.

A total of four scans were run; two where either one of \hone\ and \htwo\ was allowed to be the SM-like one, and since this mostly gave points with \honesm, two other scans were run including a bias towards \htwosm. In both cases did we use one scan focusing on the naturalness limit and one wider scan. The parameter ranges for all scans are given in table~\ref{tab:scans}, where the reduced range focuses on the naturalness
limit and, since the couplings relevant to e.g.,\ $H_i\to A_1A_1$ decays
depend on \lam\, this is the most promising part of parameter space to look for
these channels. All the scan results are then combined into two samples, one with \honesm\ and one with \htwosm.

For all samples we require \mhsm\ to lie between 122 and 129 GeV. The best experimental values of the Higgs mass are 125.03 GeV from CMS\cite{CMS-PAS-HIG-14-009} and 125.36 GeV from ATLAS\cite{Aad:2014aba} with uncertainties of the scale of a fraction of a GeV, however, to allow for potentially large theoretical uncertainties, we allow a significantly larger range. Note, though, that the benchmark points of section~\ref{benchmarks} are within the experimental limits.

\begin{table}[tbp]
\begin{center}
\begin{tabular}{|c|c|c|}
\hline
 Parameter & Extended range & Reduced range  \\
\hline
\mzero\,(GeV) & 200 -- 2000 & 200 -- 2000 \\
\mhalf\,(GeV)  & 100 -- 2000 & 100 -- 1000 \\
\azero\,(GeV)  &  $-5000$ -- 5000 & $-3000$ -- 3000 \\
\mueff\,(GeV) & 50 -- 1000 & 100 -- 200 \\
\tanb & 1 -- 30 & 1 -- 6 \\
\lam  & 0.01 -- 0.7 & 0.4 -- 0.7  \\
\kap  & 0.01 -- 0.7 & 0.01 -- 0.7 \\
\alam\,(GeV) & 200 -- 2000 & 200 -- 1000 \\
$M_p$ (GeV) & 3 -- 140 & 3 -- 140 \\
\hline
\end{tabular}
\caption{Parameter ranges used in the scans. The reduced range focuses on the naturalness limit.}
\label{tab:scans}
\end{center}
\end{table}

The scans also contain a bias towards low pseudoscalar masses
(favouring $m_{A_1}\lesssim 65$ GeV, but allowing $m_{A_1}$ up to 140
GeV) and SM-like signals rates for $H_{\rm SM}$. These constraints are
implemented to ensure that the scans do not waste too much time
exploring uninteresting parts of parameter space. Conversely, they
should not exclude any points that might be of interest for further
investigation.

Though similar scans were also performed in our previous paper\cite{Bomark:2014gya}, there are a number of important updates. Apart from updated versions of some of the software (most importantly NMSSMTools), the most notable difference is the treatment of the constraints on signal rates for \hsm;\ this is now done using the built in constraints of NMSSMTools, which uses constraints obtained from Lilith 1.0~\cite{Lilith}. Especially important is the inclusion of signal constraints on $\hsm \to b\bar b$; it turns out that virtually all points in the naturalness limit with $\maone< \mhsm/2$ are excluded by this constraint. Given the poor measurement of this channel as compared to e.g.\ $\hsm\to\gamma\gamma$ and $\hsm\to ZZ$, this might sound surprising, but it turns out that the latter channels can afford a relatively large BR($\hsm\to\aone\aone$) by compensating with larger branching ratios for the measured channels, this however is not possible for $\hsm\to b\bar b$ since this branching ratio cannot be further increased and points with increased vector boson channels tend to have some singlet component in the \hsm\ which further reduces the production cross-section and therefore conflicts with $\hsm\to b\bar b$. In the case of \honesm\ this only confirms the picture from\cite{Bomark:2014gya} where only a wider scan yields points below $\hsm/2$. For \htwosm\ on the other hand, this makes a big difference; while our previous studies deemed $\hone \to \aone\aone$ with \htwosm\ a very promising channel, this has now changed.

\begin{figure}[tbp]
\centering
\includegraphics*[width=0.45\textwidth]{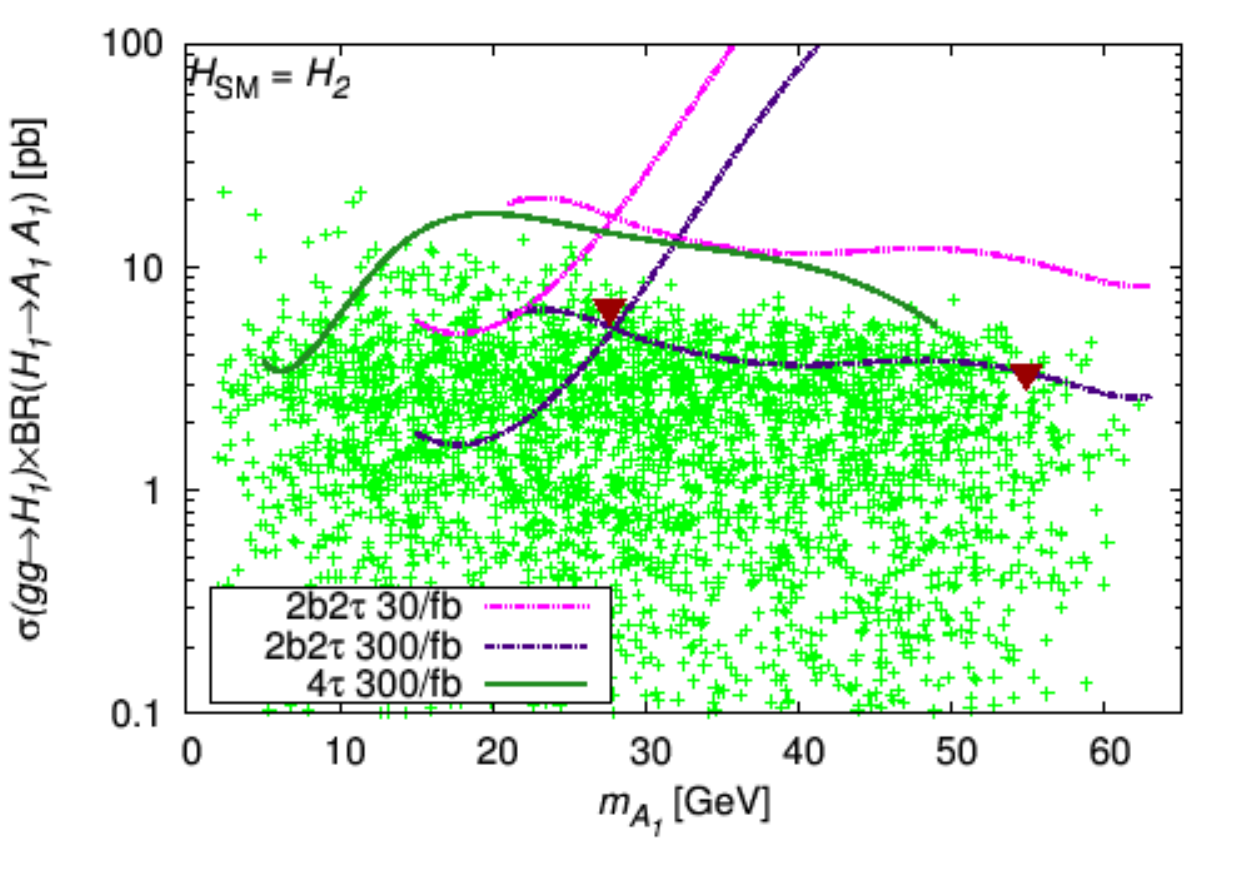}
\caption{ The sensitivity in the previously most promising channel with $gg\to H_1 \to A_1A_1$ with \htwosm, sensitivity curves are identical to the ones in figure 12(a) of \cite{Bomark:2014gya}. The sensitivity curves correspond to the one fat jet plus two $\tau$-jets (the curves at low \maone) and the two single
b-jets plus two $\tau$-jets (the curves at higher \maone) analyses.  For the fat jet plus two $\tau$-jets \mhone\ = 100 GeV, while for the two b-jets plus two $\tau$-jets curve \mhone\ = 125 GeV, which allows the coverage of points with large \maone. The green points comply with all constraints of our scans. 
  The red triangles mark the benchmark points defined in section~\ref{benchmarks}.}
\label{fig:GF}
\end{figure}

\begin{figure}[tbp]
\centering
\subfloat[]{%
\label{fig:-a}%
\includegraphics*[width=0.45\textwidth]{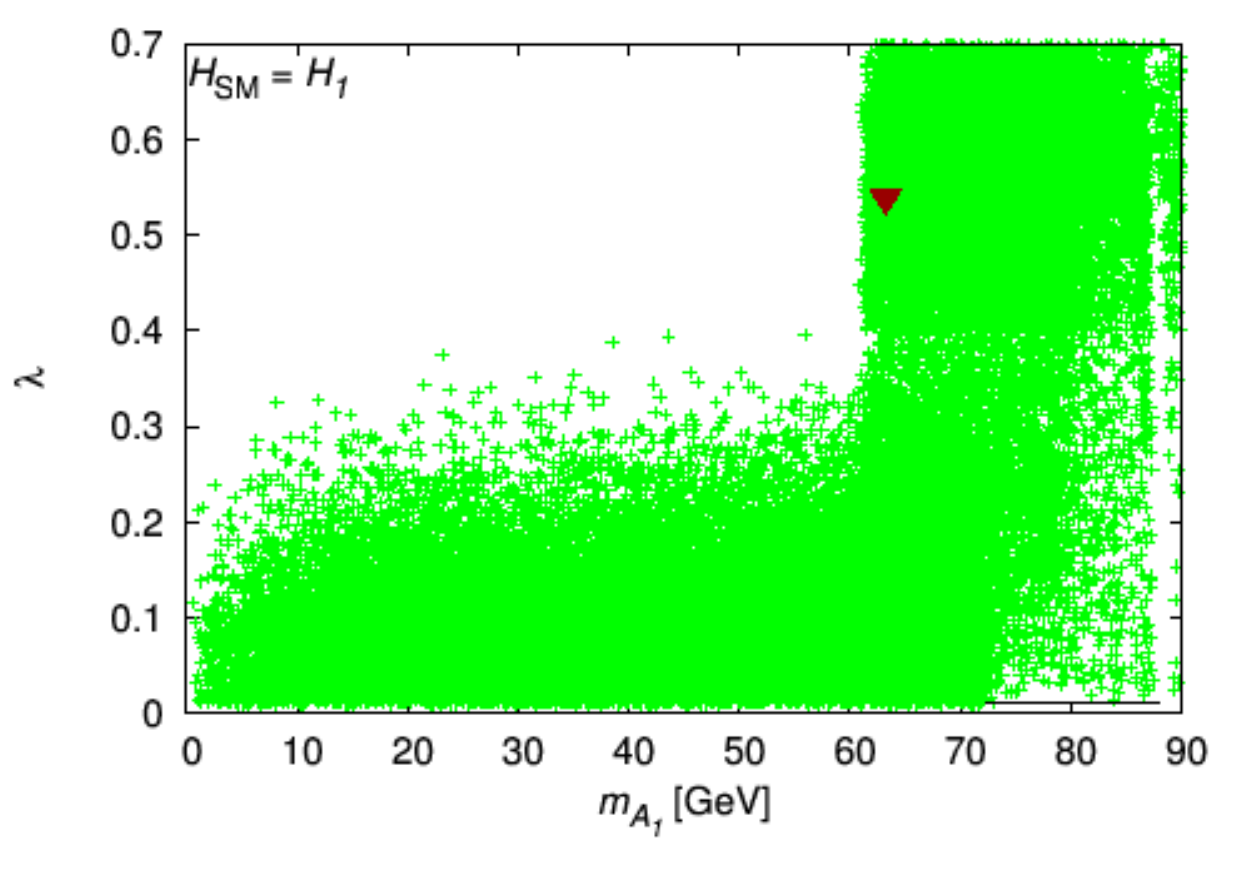}
}%
\hspace{0.5cm}%
\subfloat[]{%
\label{fig:-b}%
\includegraphics*[width=0.45\textwidth]{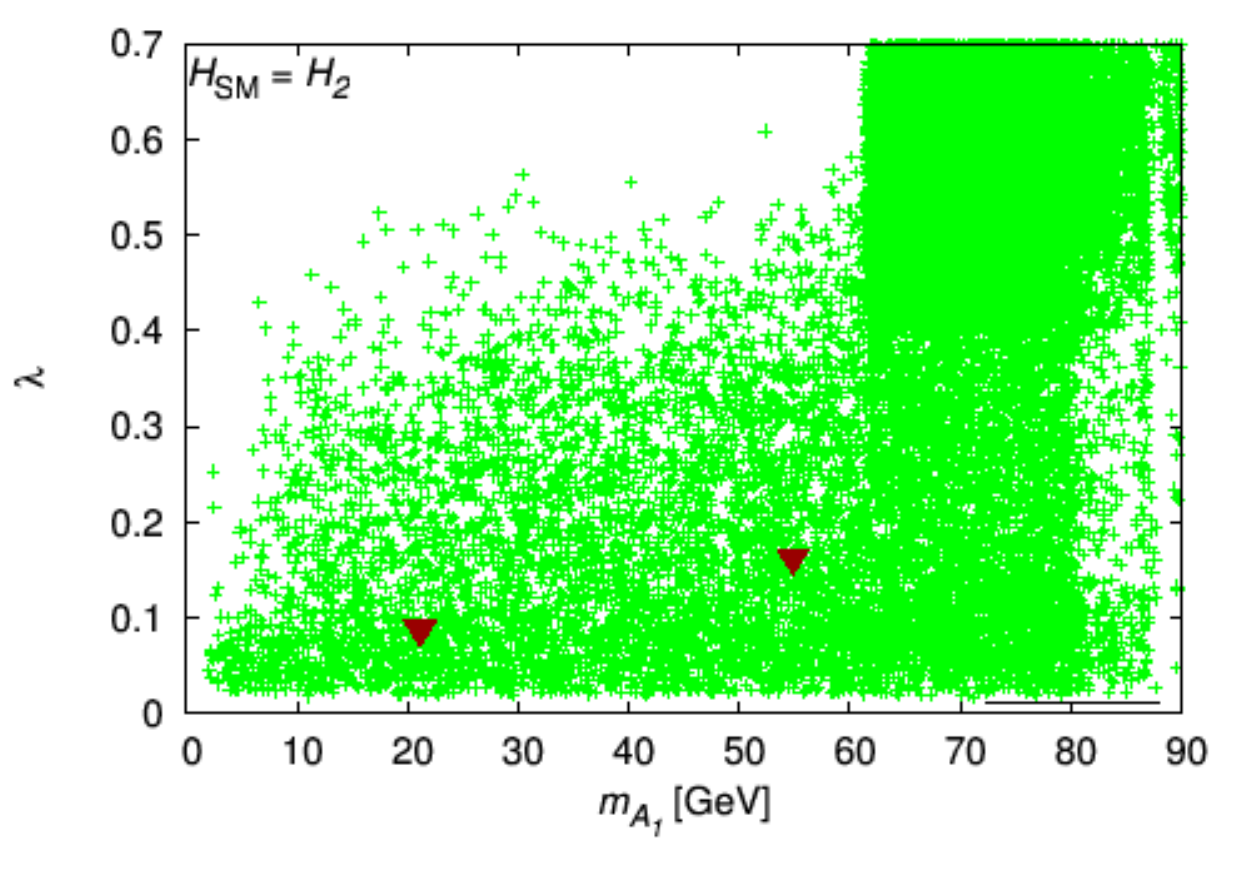}
}
\caption{The parameters \lam\ versus $m_{A_1}$. The colour code for
  the points is the same as in figure~\ref{fig:GF}. }
\label{fig:mA1lam}
\end{figure}

To illustrate the difference, we plot in figure~\ref{fig:GF} the same sensitivity curves as in figure 12(a) of\cite{Bomark:2014gya} but with our new scans instead. It is clear that, while the earlier scan found good chances of discovery already at 30/fb the prospects are now looking far dimmer. If we look at \lam\ as a function of \maone\ as is done in figure~\ref{fig:mA1lam}, we see clearly that it is the high \lam\ region that is unreachable for $\maone< \mhsm/2$, in this region the precence of the $\hsm\to \aone\aone$ channel suppresses the signal rates of \hsm, especially for large \lam\ (the relevant $\hsm\aone\aone$ coupling contains a factor $\lam^2$); for the vector boson channels this can be somewhat compensated by giving a somewhat larger singlet component to the \hsm, this reduces the production cross section but also the width of competing decay channels hence increasing the signal rate for the vector boson channels; however, this cannot be done simultaneously for $\hsm\to b\bar b$. With these updated constraints BR($\hsm \to\aone\aone $) never reaches more than just above 20\%.

It is also worth mentioning that compared to the scans of\cite{Bomark:2014gya}, we here use somewhat different parameters, especially, we use the singlet pseudoscalar mass term, $M_p$, as input (to improve efficiency) and all input is at the EW scale (a necessity if $M_p$ is to actually correspond to \maone). We also use the full NNLO contributions to the scalar masses in NMSSMTools, in contrast to what was done in\cite{Bomark:2014gya}. This raises the \mhsm\ especially for large stop mixing and mass, and might therefore further favour points with large \mhsm\ contributions from the stop sector rather than the NMSSM specific contributions of the naturalness limit.

Finally, in both figure~\ref{fig:mA1lam}(a) and (b), there is a depletion of points above 70--80 GeV: this is due to the bias towards low \maone\ included in the scans.

Given the requirement of SM-like signal rates for \hsm,\ it is natural that its reduced coupling to vector bosons will always be large (close to 1). The reduced coupling of $H_S$ will be smaller but can, due to mixing, still be significant, however the reduced coupling of $H_3$ always turns out very small.

To understand this let us look at the couplings of interest: they both have a factor\cite{Ellwanger:2009dp}
\begin{equation}\label{eq:VVHCoup}
(v_dS_{31}+v_uS_{32}),
\end{equation}
where $S_{31}$ and $S_{32}$ are elements of the neutral scalar mixing matrix, defined as $H_i=S_{ij}H_j^{\rm Weak}$ with $H_j^{\rm Weak}=(\Re(H_d),\Re(H_u),\Re(S))$. We know that $v_u/v_d=\tan\beta$ by definition and, due to the structure of the mixing matrices, it turns out that $S_{31}/S_{32}\approx -\tan \beta$. Hence the factor (\ref{eq:VVHCoup}) becomes small; typically, the reduced coupling tends to be $\approx 0.01$ or smaller and, since it comes squared in the production cross sections for both VBF and HS, we get a suppression factor of at least $10^{-4}$, rendering these channels useless for $H_3$. Note that the cancellation mentioned above gives a much stronger suppression than naively expected from the fact that \hsm\ has to have SM-like couplings combined with the sum-rule requiring the sum of the squares of the vector boson couplings for all three scalars to be unity.

%
%
%

\section{\label{analysis} Event analyses}
Since the vector boson couplings to $H_3$ are always very small we do not consider channels including $H_3$ production here, they do not have big enough cross sections to be of any interest. Furthermore,
since  the channels $H_{1,2}\to A_1 Z$ were shown in\cite{Bomark:2014gya} to be very difficult, we here focus only on the $H_{1,2}\to A_1A_1$ channels. Given the lower cross sections for VBF and HS production as compared to GF, none of the other channels carry promise for detection.

To estimate the sensitivity in the channels of interest we use
MadGraph5\_aMC$@$NLO\cite{Alwall:2014hca}, employing default parton
distribution functions and factorisation and renormalisation scales,
to generate the relevant backgrounds. Hadronisation and signal
generation is then done using Pythia 8.180\cite{Sjostrand:2007gs},
while jet clustering and jet substructure studies are done using
FastJet-v3.0.6\cite{Cacciari:2011ma}.

The production cross sections for the signals are calculated using tabulated cross sections for the SM Higgs, taken from\cite{Dittmaier:2011ti}, together with reduced couplings for the relevant scalar from \nmssmtools.

For all final state objects we use the following acceptance cuts:
\begin{itemize}
\item $|\eta|<$ 2.5 for all final state objects,
\item $p_T>15$\gev\ for all final state jets ($\tau$-jets, $b$-jets and light-quark jets),
\item $p_T>10$\gev\ for all final state leptons ($e^\pm,\ \mu^\pm$),
\item $\Delta R \equiv \sqrt{(\Delta\eta)^2+(\Delta\phi)^2}>0.2$ for all $b$-quark pairs,
\item $\Delta R >0.4$ for all other pairs of final state objects,
\end{itemize}
where $p_T$, $\eta$, $\phi$ are the transverse momentum,
pseudorapidity and azimuthal angle, respectively. Both $B$-mesons and $\tau$s are decayed by Pythia and the decay products included in the jet clustering in FastJet. The identification of $b$- and $\tau$-jets is then done by tracing the jet constituents back in the event record, thereby identifying $b$- and $\tau$-jets as jets where most of the constituents are decay products from $B$-mesons and $\tau$s respectively. On top of this the event sample is rescaled by an assumed tagging efficiency of 50\% for each $b$- and $\tau$-jet. This efficiency is somewhat conservative (c.f.\ \cite{CMStautag,CMSbtag}) but this is motivated by our soft signals. No knowledge about the charge of the jets is assumed. For the $t\bar t$ background for the $WH$ channel, we assume a 1\% mistagging probability for light jets.

For the VBF backgrounds the two additional light jets are required to satisfy the following criteria:
\begin{itemize}
\item $|\eta|<$ 5,
\item $p_T>30$\gev,
\item $\Delta R >3$,
\item $M_{jj}>300$ \gev,
\end{itemize}
where $M_{jj}$ is the invariant mass of the two forward/backward
jets. (Due to numerical difficulties in producing enough statistics,
somewhat harder cuts, $M_{jj}>500$ GeV and $p_T>40$ GeV, are imposed
on the parton level production in the $4b$ final state.)  In addition,
the jets are required to be located in the opposite hemispheres (i.e.\
one positive and one negative $\eta$) and there should be no other jet
(apart from the signal objects) with $p_T>30$ GeV between (in terms of
$\eta$) the two jets.

The cross sections for the dominant backgrounds after the above acceptance cuts are applied, are given in table~\ref{tab:BackXS}. Note that for the $0j$ and $1j$ backgrounds the above VBF specific cuts are not applied, hence the much larger cross-sections, this will be compensated by smaller acceptance after hadronisation. After hadronisation and jet clustering the final VBF cuts are applied to the real jets (as opposed to the parton as done above) and then $M_{jj}>500$ GeV and $p_T>40$ GeV are required for all signal and background in the VBF channels. For the rest of the paper, in all discussions of VBF production $t\bar t$ refers to the sum of $0j+t\bar t$, $1j+t\bar t$ and $2j+t\bar t$ and $2b2\tau$ refers to the sum of $0j+2b2\tau$, $1j+2b2\tau$  and $2j+2b2\tau$. The total background for the $2b2\tau$ channel is the sum of $t\bar t$ and $2b2\tau$. We do not include further backgrounds in the $4b$ channel since this channel is in any case inferior to  $2b2\tau$ and therefore of little interest.

\begin{table}[tbp]
\begin{center}
\begin{tabular}{|l|c|}\hline
Channel  &  Parton level cross section   \\\hline
\multicolumn{2}{|c|}{VBF}\\\hline
$2j+4b$  &  72 pb    \\
$0j+2b2\tau$  &  3.1 pb    \\
$1j+2b2\tau$  &  3.2 pb    \\
$2j+2b2\tau$  &  0.19 pb    \\
$0j+t\bar t$ &  597 pb  \\
$1j+t\bar t$ &  845 pb  \\
$2j+t\bar t$ &  80 pb  \\\hline
\multicolumn{2}{|c|}{$ZH$}\\\hline
$Z+4b$   &  0.31 pb   \\
$Z+2b2\tau$  &  2.7 fb \\\hline
\multicolumn{2}{|c|}{$WH$}\\\hline
$W+4b$    &   36 fb     \\
$t\bar tb\bar b$  &  4.0 pb   \\
$t\bar t$  &  597 pb    \\\hline
\end{tabular}
\caption{Background cross sections for the dominant backgrounds at
  parton level as calculated by MadGraph.}\label{tab:BackXS}
\end{center}
\end{table}%

In order to optimise the sensitivity to boosted $A_1$s we employ the
jet substructure method of\cite{Butterworth:2008iy} (see
also\cite{Bomark:2014gya} for further details). This gives us fat jets
that we assume to originate from an $A_1$ decaying into a $b\bar b$
pair. To avoid contamination from single $b$-jets, we require all fat
jets to have $p_T>30$ GeV and invariant mass $>12$ GeV.

For each event with the proper final state we look for two $A_1$
candidates (either one fat jet, two $b$-jets or two $\tau$-jets
depending on which channel we are looking at) and compare their
respective invariant mass: if they are within 15 GeV of each other we
accept that event and if, in addition, the combined invariant mass of
the two candidates is within $125\pm30$ GeV, it is accepted as an
event where an $H_{\rm SM}$ was produced and decayed into two
$A_1$s. In all the channels we perform two analyses in parallel: one with
the jet substructure method, where only fat jets and no single
$b$-jets are used, and one where no jet substructure is exploited but
only single $b$-jets are used. For the four $b$-jet final state we check
all combinations of $b$-jet pairs and accept the first one with both invariant masses within 15 GeV of each other.

To obtain the sensitivity for a given $m_{A_1}$ we then count the events where the $A_1$ candidates masses are within 15 GeV of the mass $m_{A_1}$ and can then calculate --- for a given luminosity, $\mathcal L$ --- how large a signal cross section is needed to obtain a significance $S/\sqrt B > 5$ where $S$ is the number of signal events and $B$ is the background. In all channels we require at least 10 events in order to claim discovery, so if  $S/\sqrt B > 5$ is fulfilled for $S<10$, we instead use $S>10$ as the limiting sensitivity. This means that, in channels with very low background, the sensitivity is $\propto \mathcal L$ rather than $\propto \sqrt{\mathcal L}$ as is the case for $S/\sqrt B$.

We will be studying the VBF and HS channels in turn, splitting the latter into the $ZH$ and $WH$ modes. For the sensitivity curves in the forthcoming sections, we limit ourselves to scalar masses up to 175 GeV, there is no problem in principle to employ these analyses at higher masses, they would actually gain some sensitivity with harder objects to study, but the production cross-section for the initial scalar also drops fast with increasing mass, hence prohibiting any discovery prospects.

\section{\label{VBF} VBF}

The backgrounds used in the analysis of the VBF mode
are irreducible, i.e.,\ $4b+$2 jets and $2b2\tau+$2 jets. In addition, we include the $t\bar t+$2 jets --- with both $W$s from the top quarks decaying to $\tau$s --- in the $2b2\tau$ channel. Since the latter turns out to be the dominant background in this channel, we also invoke a cut on Missing Transverse Momentum (MET) to reduce it.

This cut requires that the $p_T$ of the two $\tau$-jets combined should be larger than the total MET of the event. This reduces the $t\bar t+$2 jets background by a factor 2-3 while leaving both signal and irreducible backgrounds virtually intact. Note, however, that the MET here is simply the sum of the momentum of all invisible particles (i.e.,\ neutrinos), a full detector simulation with mis-measured/missing jets, pile-up etc.\ would be necessary to fully determine the true effectiveness of this cut. This latter cut, though effective, is, however, not crucial for the usefulness of this analysis.

\begin{figure}[tbp]
\centering
\includegraphics*[width=0.45\textwidth]{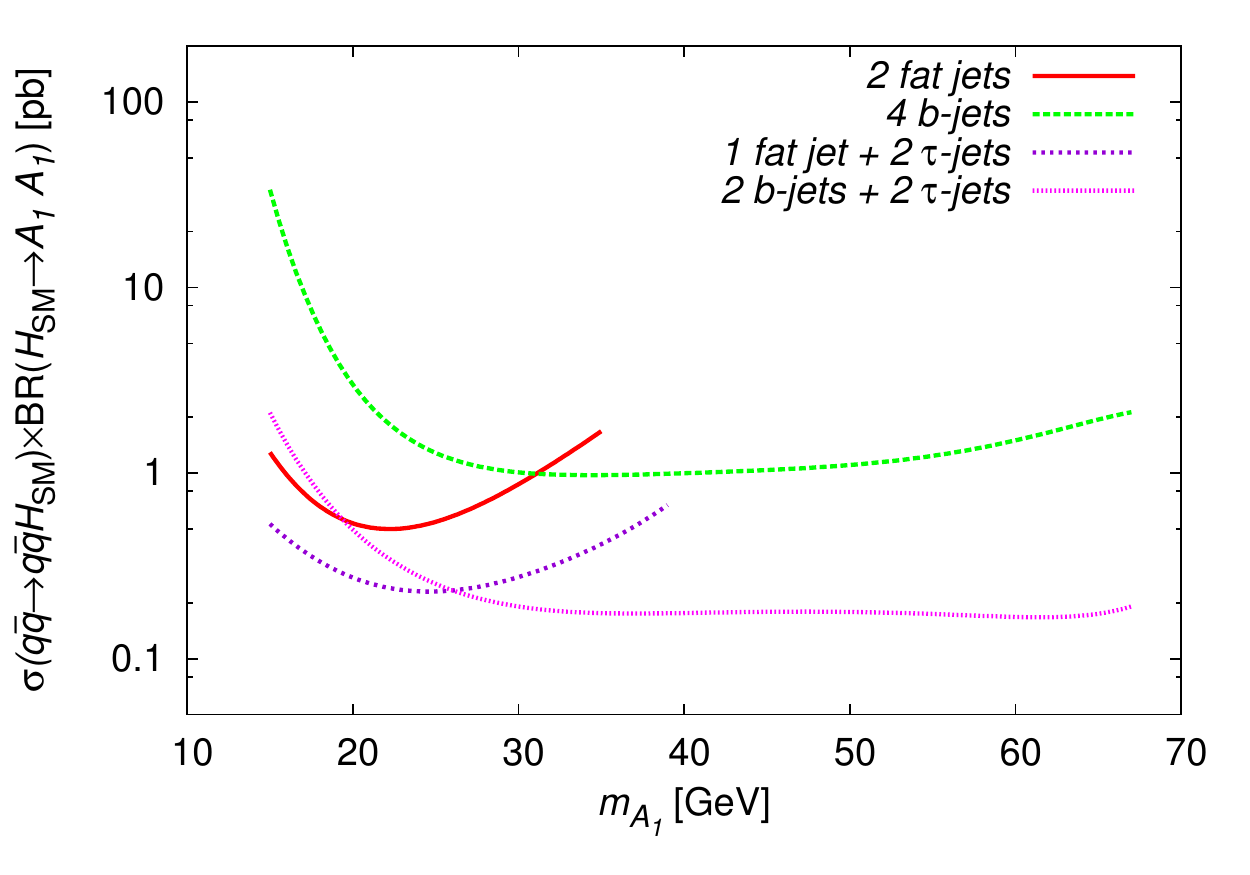}
\caption{Sensitivity for the $4b$ and $2b2\tau$ channels in VBF production. All curves assume 3000/fb of
integrated luminosity and use $m_{H_i}=125$ GeV and the
corresponding cut on the total final state invariant mass.}
\label{fig:SensVBF}
\end{figure}

Figure~\ref{fig:SensVBF} shows the discovery reach in the interesting
channels for 3000/fb of integrated luminosity and using the overall
constraint that the total four-body invariant mass should be
125$\pm30$ GeV. It is clear that the $2b2\tau$ channels are the most
promising ones and hence we will not consider the $4b$ channels in the
following.  All sensitivity curves have been rescaled by a factor 1/0.9
for each $b\bar b$ pair in the final state and a factor 1/0.1 for each
$\tau\tau$ pair in the final state, to allow for a direct comparison
of the sensitivity to $\sigma(q\bar q\to q\bar q H_i)\times {\rm
  BR}(H_i\to A_1A_1)$.  As expected, we see in
figure~\ref{fig:SensVBF} that the jet substructure analysis only works
well for rather low $m_{A_1}$.

\subsection{$\boldsymbol{H_1}$ SM-like}

\begin{figure}[tbp]
\centering
\subfloat[]{%
\label{fig:-a}%
\includegraphics*[width=0.45\textwidth]{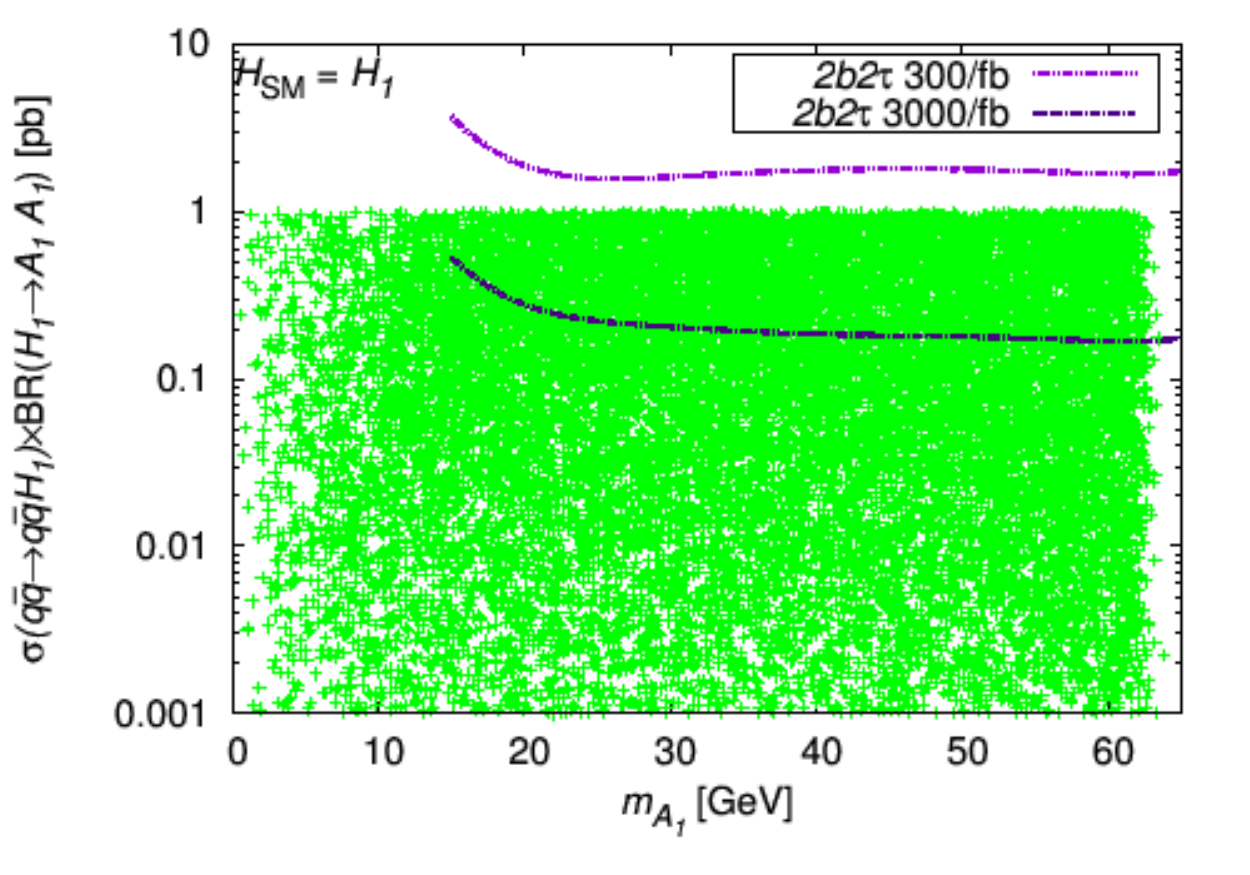}
}%
\hspace{0.5cm}%
\subfloat[]{%
\label{fig:-b}%
\includegraphics*[width=0.45\textwidth]{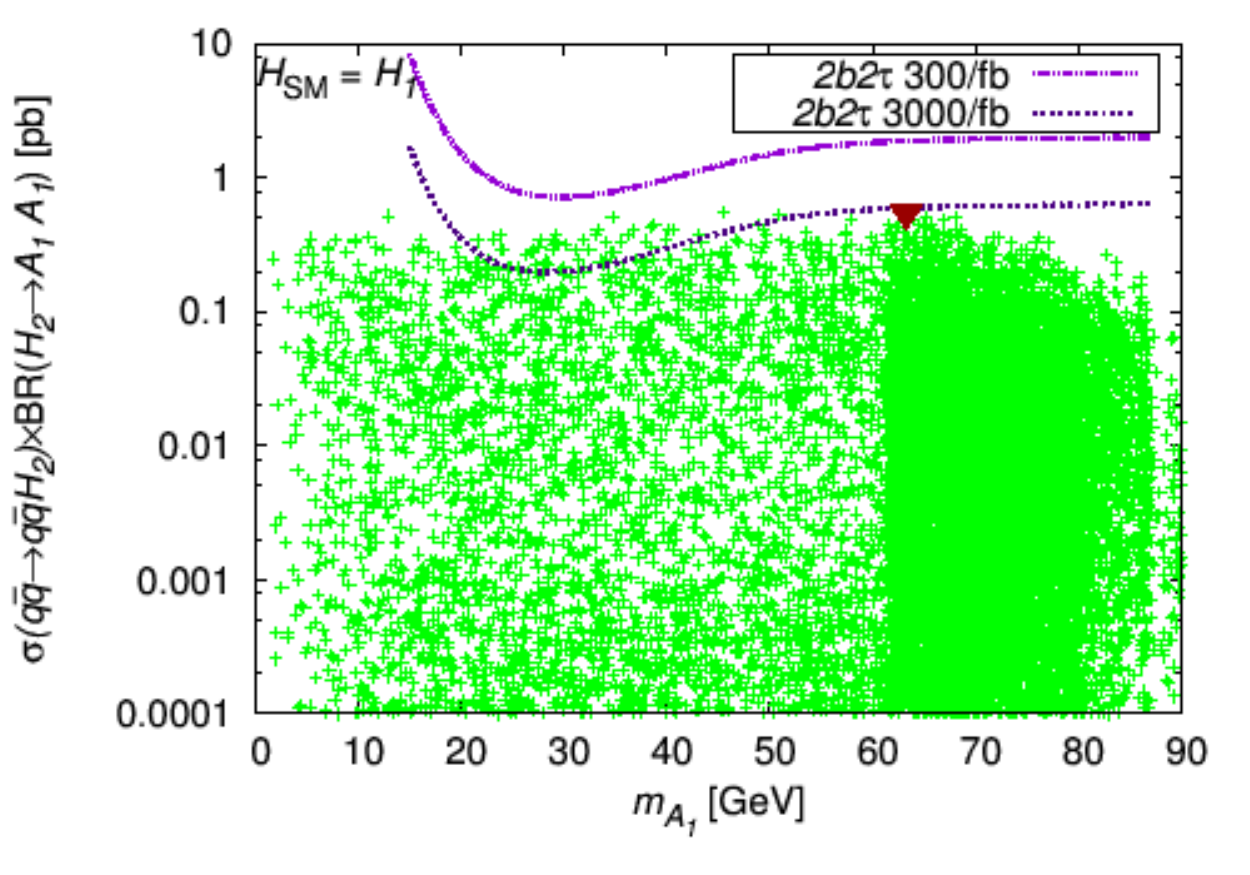}
}
\caption{LHC reach in $H_1\to A_1A_1$ (left) and $H_2\to A_1A_1$
  (right) for $H_1=H_{\rm SM}$ in the VBF channel. In panel (a) the sensitivity curves use \mhone\ =125 GeV, while (b) uses  $m_{H_2}=175$ GeV and both uses the best of the 2 $b$-jets and fat jet analyses. The colour code for
  the points is the same as in figure~\ref{fig:GF}.}
\label{fig:h1VBF}
\end{figure}

To illustrate the reach of the LHC in these channels, in
figure~\ref{fig:h1VBF} we compare  the sensitivity with the points from our scans
when $H_1=H_{\rm SM}$. In figure~\ref{fig:h1VBF}(a), we show
sensitivity curves using a 125 GeV Higgs (and employing the
corresponding constraint), always using the best out of the 2 $b$-jets
and fat jet analyses. As is clearly seen, the prospects for $H_1\to
A_1A_1$ will always be rather limited due to the small number of
points in the interesting region of parameter space. It is clear that the points never reach above 1 pb, this is a direct effect of the 
constraints on the signal rates of \hsm, higher BR$(\hsm\to A_1A_1)$ would necessarily suppress other channels unacceptably much.
This means that we need 3000/fb to discover something and our best chance would be the low mass region where the jet substructure methods can improve sensitivity.

As regards the $H_2\to A_1A_1$ channel, it is somewhat easier to find acceptable points just above the $\maone < \mhone/2$ threshold, however, they do not reach much higher in terms of rates.
This can be seen in figure~\ref{fig:h1VBF}(b), where the prospects for $H_2\to
A_1A_1$ are illustrated. Figure~\ref{fig:h1VBF}(b) uses sensitivity
curves for $m_{H_2}=175$ GeV in order to cover the whole interesting
parameter space. Also, here do we always employ the analysis (with or
without jet substructure) with the best sensitivity.
The smallness of the rates also above the kinematic threshold is a consequence of the requirement that \hsm\ has to have very SM-like couplings to meet the signal rate constraints, and hence \htwo\  has to be very singlet-like and hence have very small production rates even if the $\hone\to A_1A_1$ channel is kinematically closed. In the end this means that our best chance also in this channel seems to be the light pseudoscalars where jet substucture is useful.

\subsection{$\boldsymbol{H_2}$ SM-like}
\begin{figure}[tbp]
\centering
\subfloat[]{%
\label{fig:-a}%
\includegraphics*[width=0.45\textwidth]{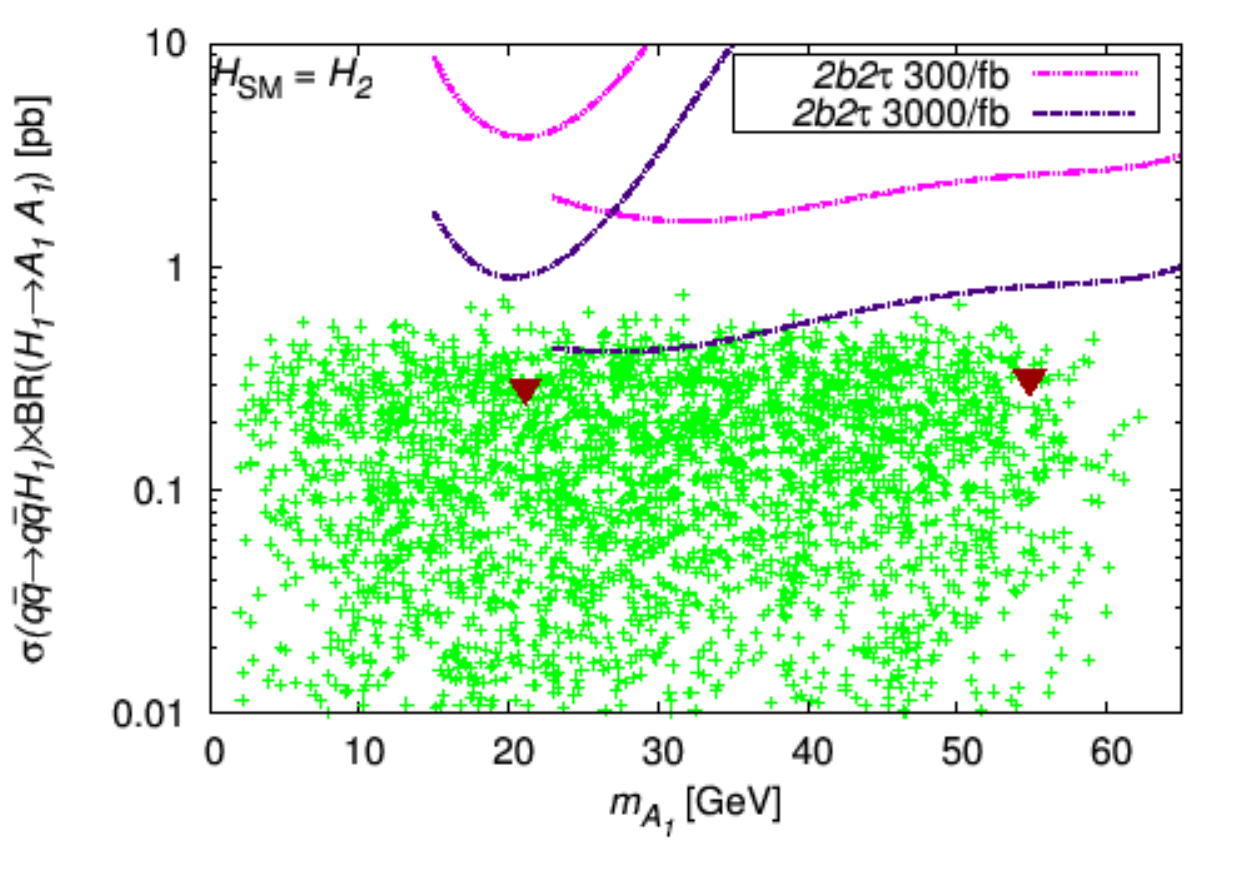}
}%
\hspace{0.5cm}%
\subfloat[]{%
\label{fig:-b}%
\includegraphics*[width=0.45\textwidth]{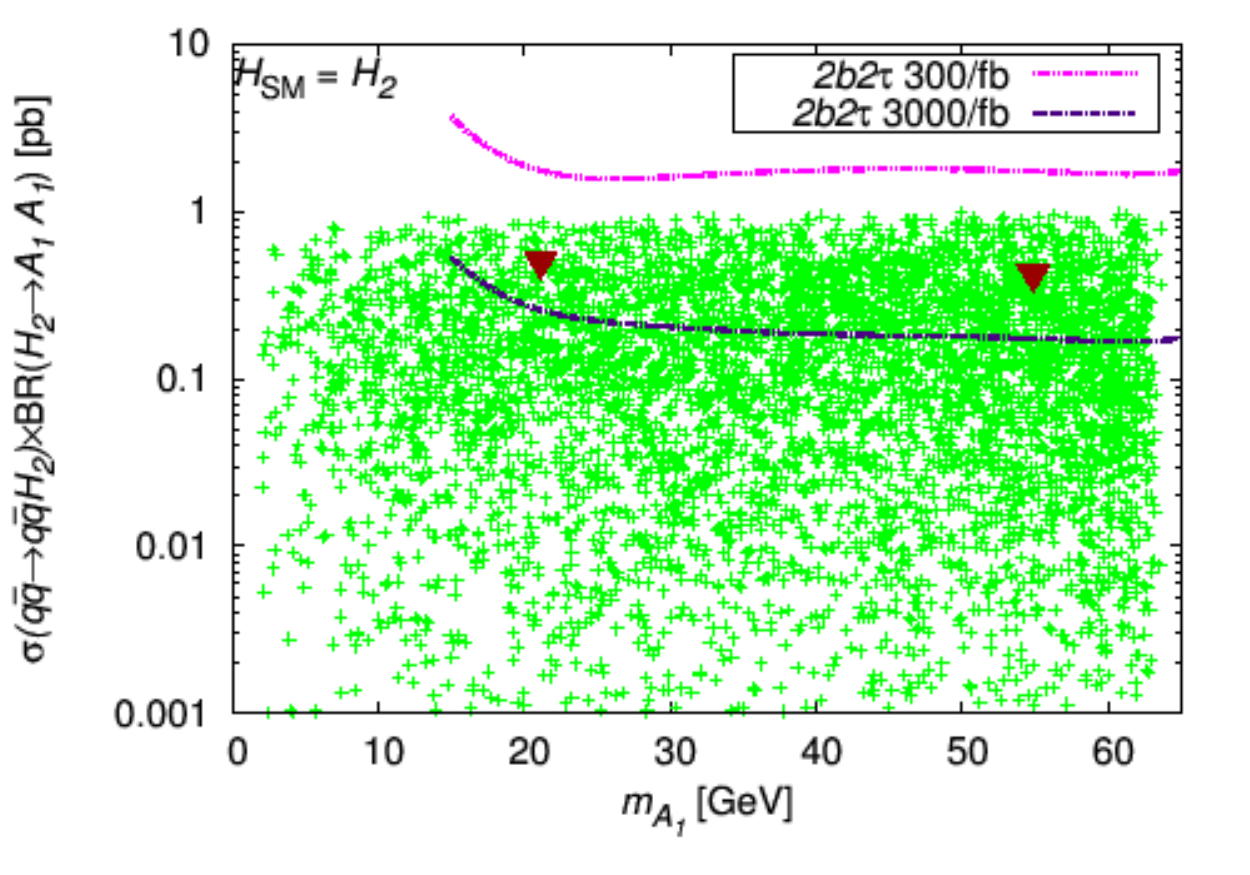}
}
\caption{LHC reach in $H_1\to A_1A_1$ (left) and $H_2\to A_1A_1$
  (right) for $H_2=H_{\rm SM}$ in the VBF channel. The sensitivity curves in panel (a) correspond to the one fat jet (the curves at low \maone\ using \mhone\ = 100 GeV) and the two single b-jets (the curves at higher \maone\ using \mhone\ = 125 GeV) analyses. In panel (b) \mhtwo\ = 125 GeV is used along with the best of the 2 $b$-jets and fat jet analyses. The colour code for
  the points is the same as in figure~\ref{fig:GF}.}
\label{fig:h2VBF}
\end{figure}

As mentioned before, after the inclusion of especially the signal rate constraint on $\hsm\to b\bar b$, the case $H_2=H_{\rm SM}$ does not differ so much from $H_1=H_{\rm SM}$.
In figure~\ref{fig:h2VBF}(a) we show the
sensitivity for $H_1\to A_1A_1$, both with jet substructure using
$m_{H_1}=100$ GeV and without jet substructure using $m_{H_1}=125$
GeV, but without constraining the four-body invariant mass. We use two
separate curves since they use different $m_{H_1}$. The use of \mhone\ = 100 GeV is motivated by the typical values of \mhone\ while 125 GeV is needed to cover the whole range of \maone\ , the \maone\ = 100 GeV curve would be kinematically cut off at \maone\ = 50 GeV. Also in this channel we will need 3000/fb to see something.

From figure \ref{fig:h2VBF}(a) it even looks like the jet substructure does worse than the simpler 2 $b$-jet analysis, though one should remember that these curves uses different \mhone; the lower \mhone\ of the fat jet analysis means decreased sensitivity. For the $4b$ final state this can be
compensated by an improved sensitivity due to the fat jet analysis,
however, as can be seen in figure~\ref{fig:SensVBF}, the low mass
improvement with the fat jet analysis is not as great for $2b2\tau$ as
it is for $4b$, leading to comparatively poor sensitivity to
$\maone\approx 20$ GeV in figure~\ref{fig:h2VBF}(a). 

In figure~\ref{fig:h2VBF}(b) we show the sensitivity in the $H_2\to
A_1A_1$ channel using $m_{H_2}=125$ GeV and requiring the four-body
invariant mass to be $125\pm 30$ GeV. For all points along the
sensitivity curves we use the analysis that gives the best sensitivity
(with or without jet substructure). As expected, this is very similar to figure~\ref{fig:h1VBF}(a) with a clear upper limit for the rate at 1 pb stemming from the signal rate requirements on \hsm.

\section{\label{ZH} Higgs-strahlung via $\boldsymbol{ZH}$}

\begin{figure}[tbp]
\centering
\includegraphics*[width=0.45\textwidth]{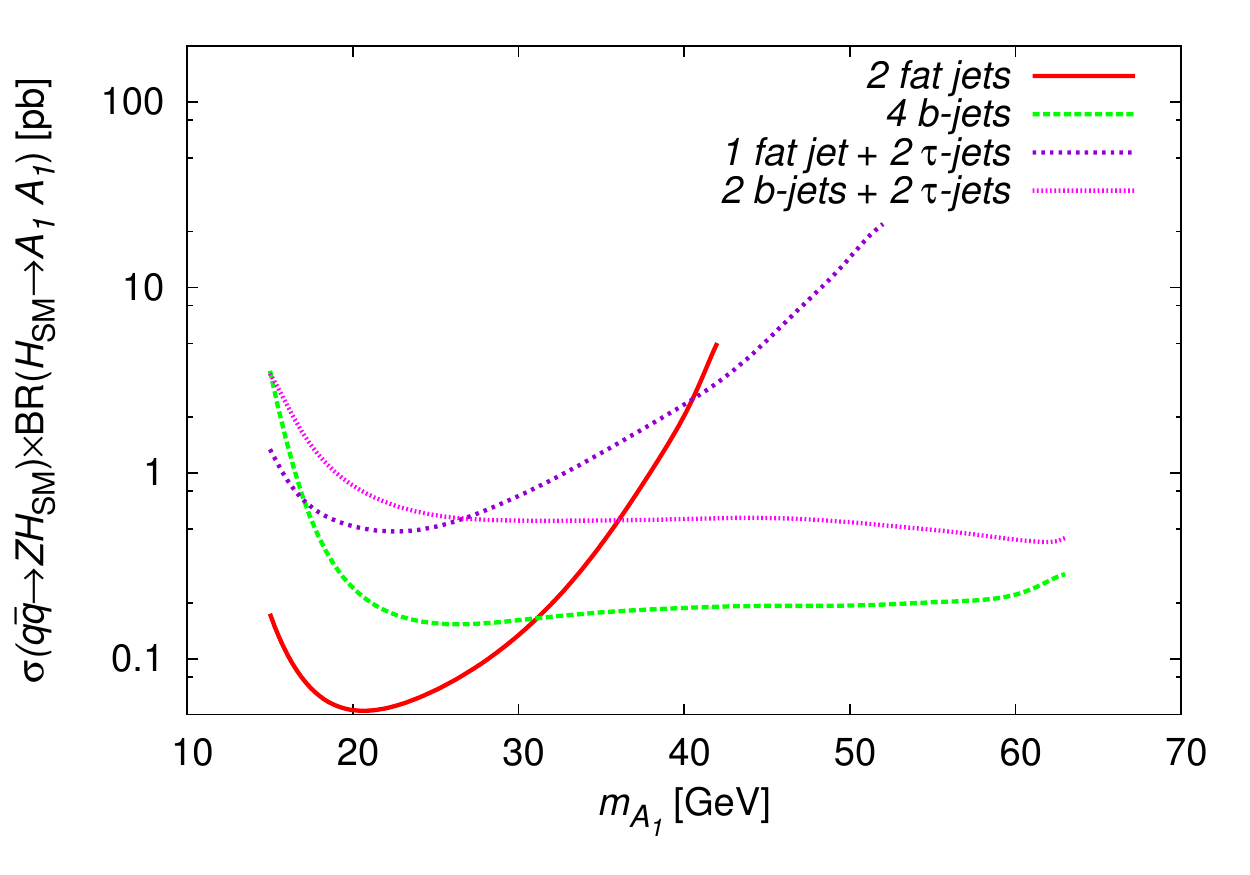}
\caption{Sensitivity for HS off a $Z$ boson. All curves assume 3000/fb of
integrated luminosity and use $m_{H_i}=125$ GeV and the
corresponding cut on the total final state invariant mass.}
\label{fig:SensZH}
\end{figure}

With an additional $Z$ boson, triggering and background suppression is
much improved. To extract the signal we only use leptonically decaying
$Z$ bosons. This means, that in addition to acceptances, we require
one di-lepton pair with invariant mass $90\pm 10$ GeV.  This is very
powerful in suppressing the backgrounds, however, the small leptonic
BR of the $Z$ together with the small production cross sections in the
$ZH$ channel, means that one will struggle to get a large enough signal.
As one would then expect, the best final state to look for is not
$2b2\tau$ as was used before, but rather $4b$ which gives the highest
signal rate. This is clearly seen in figure~\ref{fig:SensZH}, where
the sensitivity in the various channels are shown for 3000/fb of
integrated luminosity. All channels use $m_{H_i}=125$ GeV and the
corresponding cut on the total final state invariant mass.

\subsection{$\boldsymbol{H_1}$ SM-like}
\begin{figure}[tbp]
\centering
\subfloat[]{%
\label{fig:-a}%
\includegraphics*[width=0.45\textwidth]{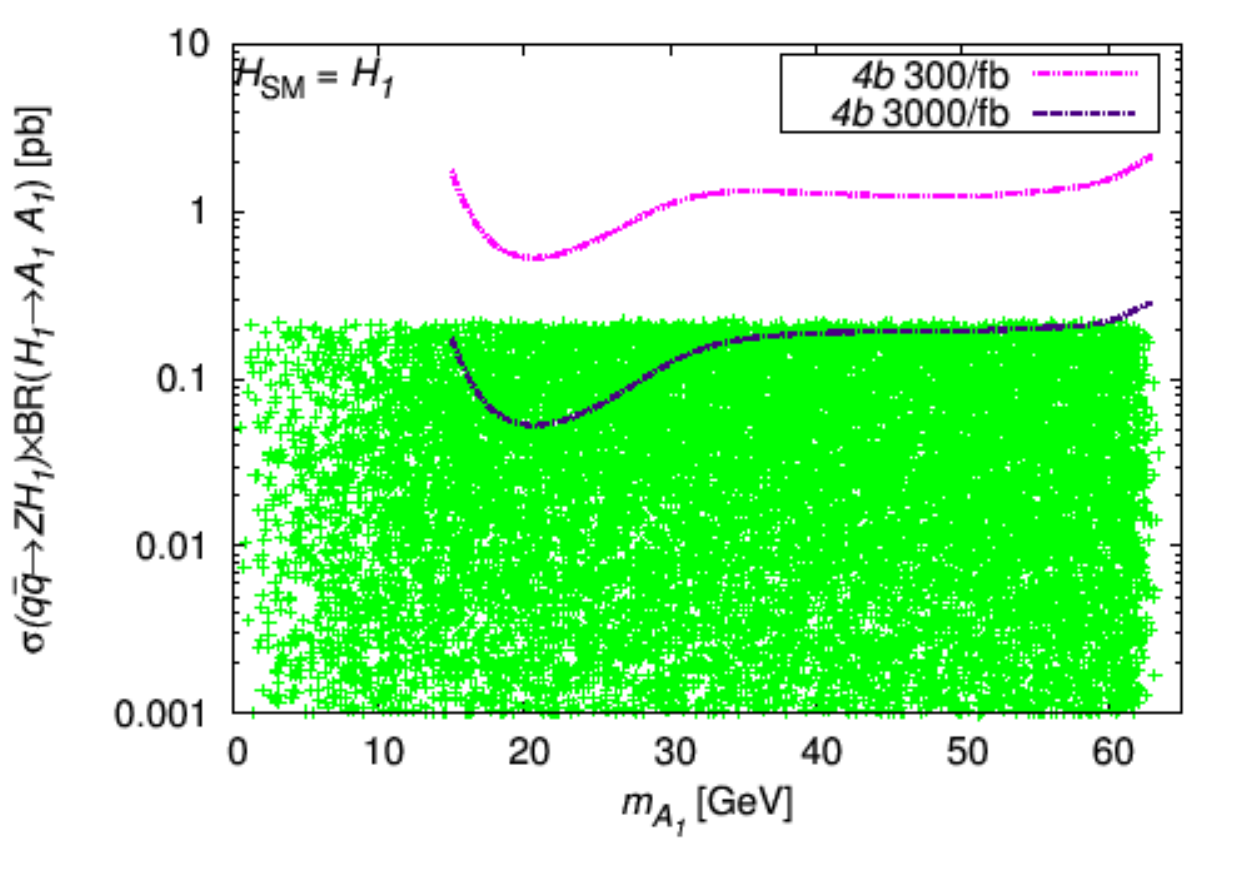}
}%
\hspace{0.5cm}%
\subfloat[]{%
\label{fig:-b}%
\includegraphics*[width=0.45\textwidth]{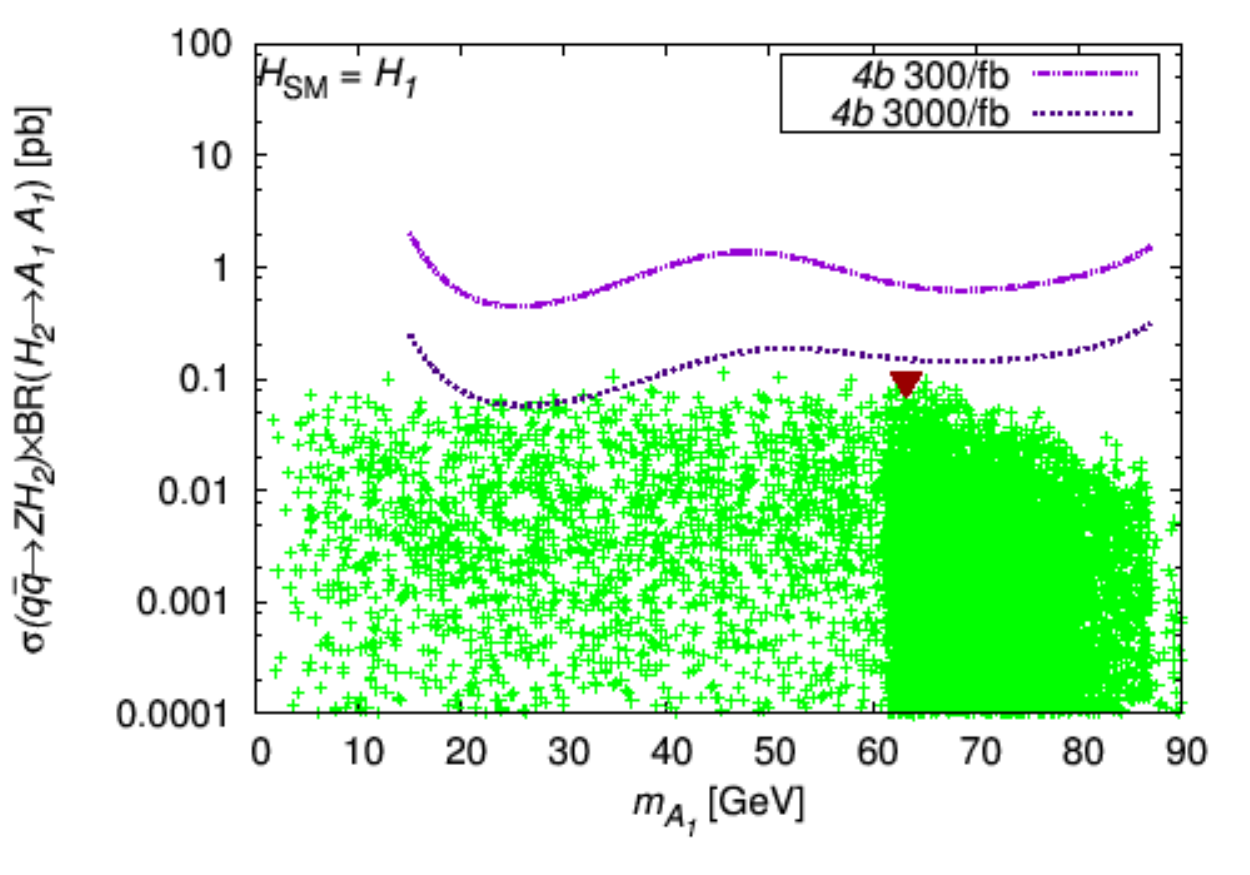}
}
\caption{LHC reach in $H_1\to A_1A_1$ (left) and $H_2\to A_1A_1$
  (right) for $H_1=H_{\rm SM}$ in the $ZH$ channel. In panel (a) the sensitivity curves use \mhone\ =125 GeV, while (b) uses  $m_{H_2}=175$ GeV and both uses the best of the 4 $b$-jets and fat jets analyses. The colour code
  for the points is the same as in figure~\ref{fig:GF}.}
\label{fig:h1ZH}
\end{figure}

Similarly to the VBF case, the SM-like $H_1$ scenario is difficult
with respect to detection. In figure~\ref{fig:h1ZH}(a) we show the
sensitivity in the $H_1\to A_1A_1$ channel, given $m_{H_1}=125$ GeV
(with a corresponding cut on the four-body invariant mass), as compared
to the acceptable parameter points. We notice that the prospects are
even dimmer than in the VBF case, basically our only hope is the jet substructure methods that gives a rather large improvement for low \maone, for the $4b$ final state used here this improvement can be significantly larger than for the $2b2\tau$ used in the VBF channels. As usual only the analysis with the
best sensitivity is used in each point of the curves.

Moving on to $H_2\to A_1A_1$, we see in figure~\ref{fig:h1ZH}(b) that
there is only marginal hope for detection even at 3000/fb. The curves
here correspond to $m_{H_2}=175$ GeV, with no cut on the total final
state invariant mass and always using the most efficient of the single $b$-jets/fat jets analyses. The main reason for the difficulties in this channel is clearly the small signal rates, but also the absence of a cut on the overall invariant mass impacts negatively on the sensitivity.


\subsection{$\boldsymbol{H_2}$ SM-like}
\begin{figure}[tbp]
\centering
\subfloat[]{%
\label{fig:-a}%
\includegraphics*[width=0.45\textwidth]{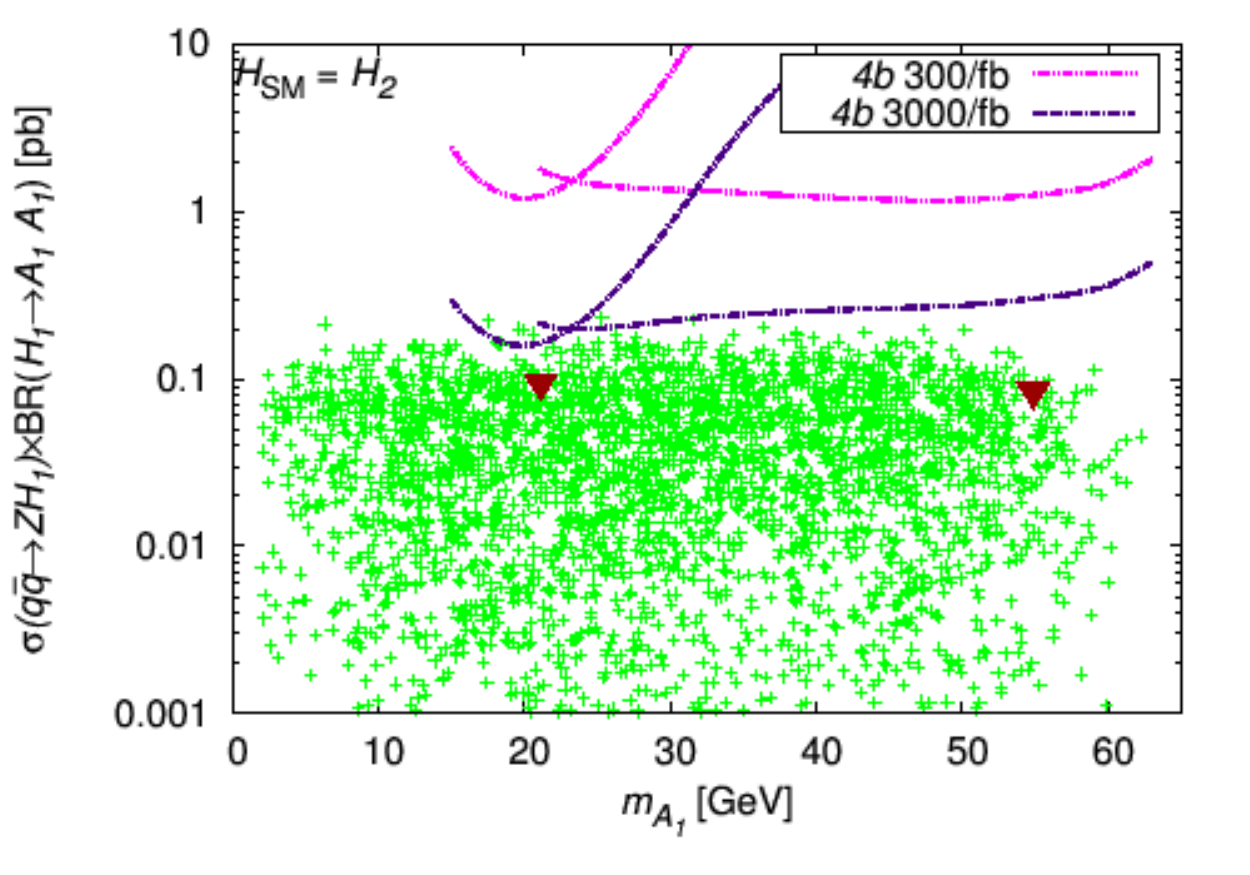}
}%
\hspace{0.5cm}%
\subfloat[]{%
\label{fig:-b}%
\includegraphics*[width=0.45\textwidth]{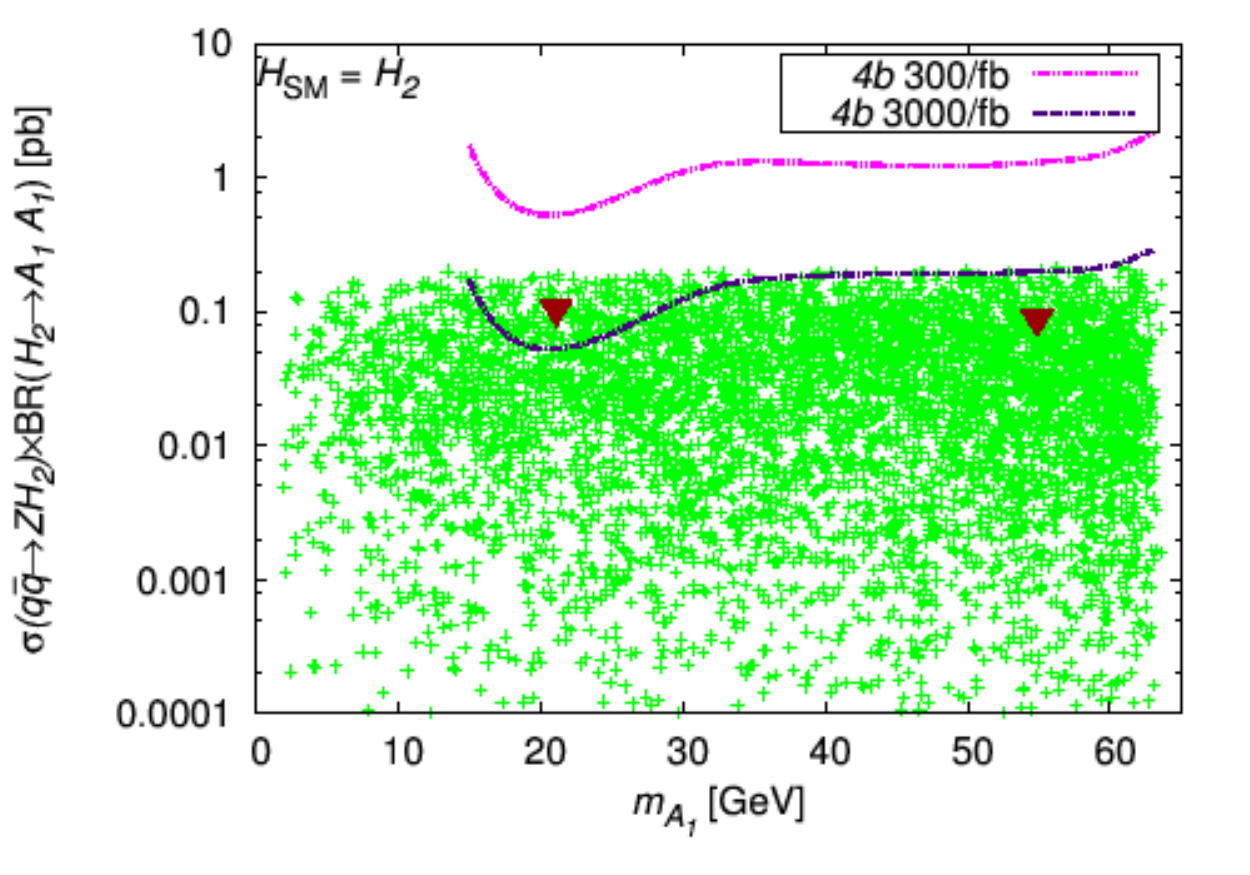}
}
\caption{LHC reach in $H_1\to A_1A_1$ (left) and $H_2\to A_1A_1$
  (right) for $H_2=H_{\rm SM}$ in the $ZH$ channel. The sensitivity curves in panel (a) correspond to the two fat jets (the curves at low \maone\ using \mhone\ = 100 GeV) and the four single b-jets (the curves at higher \maone\ using \mhone\ = 125 GeV) analyses. In panel (b) \mhtwo\ = 125 GeV is used along with the best of the 4 $b$-jets and fat jets analyses. The colour code
  for the points is the same as in figure~\ref{fig:GF}.}
\label{fig:h2ZH}
\end{figure}

With the lower sensitivity as compared to VBF production, the $H_2=H_{\rm SM}$ is also rather difficult. As can be seen for $H_1\to A_1A_1$ in
figure~\ref{fig:h2ZH}(a) --- where the sensitivity curves use
$m_{H_1}=100$ GeV with jet substructure, and $m_{H_1}=125$ GeV with
single $b$-jets (no constraint on overall invariant mass),
respectively ---  even 3000/fb is probably not sufficient for any discovery.

In figure~\ref{fig:h2ZH}(b) we show the reach for $H_2\to A_1A_1$, the
sensitivity curves correspond to $m_{H_2}= 125$ GeV and the
corresponding constrain on the four-body invariant mass, always
showing the most sensitive analysis. As before, this is basically identical to figure~\ref{fig:h1ZH}(a), with only marginal discoverability for low masses where jet substructure may be useful.

\section{\label{WH} Higgs-strahlung via $\boldsymbol{WH}$}

In general $WH$ production will have higher cross sections than $ZH$
production and, since the $W$ also has roughly three times higher
leptonic BR as compared to the $Z$, this channel will exhibit much higher
rates. In order to tag the $W$ we require exactly one isolated lepton
in the event, in addition to the signal objects. Similarly to $ZH$
production, we only look at the channel with the highest rate, i.e.,\
$W+4b$.

There has been a number of earlier studies of the $W+4b$ channel,
including parton level studies\cite{Carena:2007jk}
and\cite{Cheung:2007sva} and a full detector study
in\cite{Cao:2013gba}. While the parton level analysis
of\cite{Cheung:2007sva} arrived at significantly higher sensitivity
than we did, our results are in reasonable agreement
with\cite{Cao:2013gba}. (Note however that, although both of these
studies use four $b$-tags as well as a cut on the four-body invariant
mass, neither of them uses the requirement that both $A_1$ candidates
should have similar mass.)

In addition to the irreducible $W+4b$ background, there are
significant backgrounds from $t\bar t $ --- with two light jets from
one of the resulting $W$s being mistagged as $b$-jets --- as well as
from $t\bar tb\bar b$ events. The latter is in our studies the most
significant one, often at least one order of magnitude larger than the
irreducible background. This conclusion is in agreement
with\cite{Cheung:2007sva}, while\cite{Cao:2013gba} finds that detector
smearing pushes the $t\bar t$ background to lower invariant masses and
hence becomes a significant background also at 125 GeV.  To suppress
the $t\bar tb\bar b$ background we employ a veto against hadronically
decaying $W$s. This means an event with two light jets with $p_T>15$
GeV and combined invariant mass $= 80\pm15$ GeV is rejected as it is
likely to come from a $t\bar tb\bar b$ event with one $W$ decaying
leptonically and the other one hadronically. The events where both
$W$s decay leptonically should be suppressed by the fact that we ask
for exactly one lepton and hence reject events with two isolated
leptons (this also suppresses any $Z+4b$ backgrounds).

Since the smallness of the signal rates in both $WH$ and (even more)
in $ZH$ production is a bigger problem than background suppression,
one could consider requiring only three $b$-tags. This was the
approach of\cite{Carena:2007jk} and should yield significantly higher
signal rates, especially as one $b$-jet is often missed due to $p_T$
cuts, etc. However, this requires a much more detailed study as there
are many more contributing sources of background. It is also not clear
how to implement the invariant mass constraints as one have to assume
that sometimes the fourth $b$-jet is not selected even as a light
jet. Such considerations are therefore beyond the scope of this paper.

\subsection{$\boldsymbol{H_1}$ SM-like}
In this case higher rates as compared to $ZH$ production does mean
better discovery prospects even though the background is also larger
due to $t\bar tb\bar b$.

\begin{figure}[tbp]
\centering
\subfloat[]{%
\label{fig:-a}%
\includegraphics*[width=0.45\textwidth]{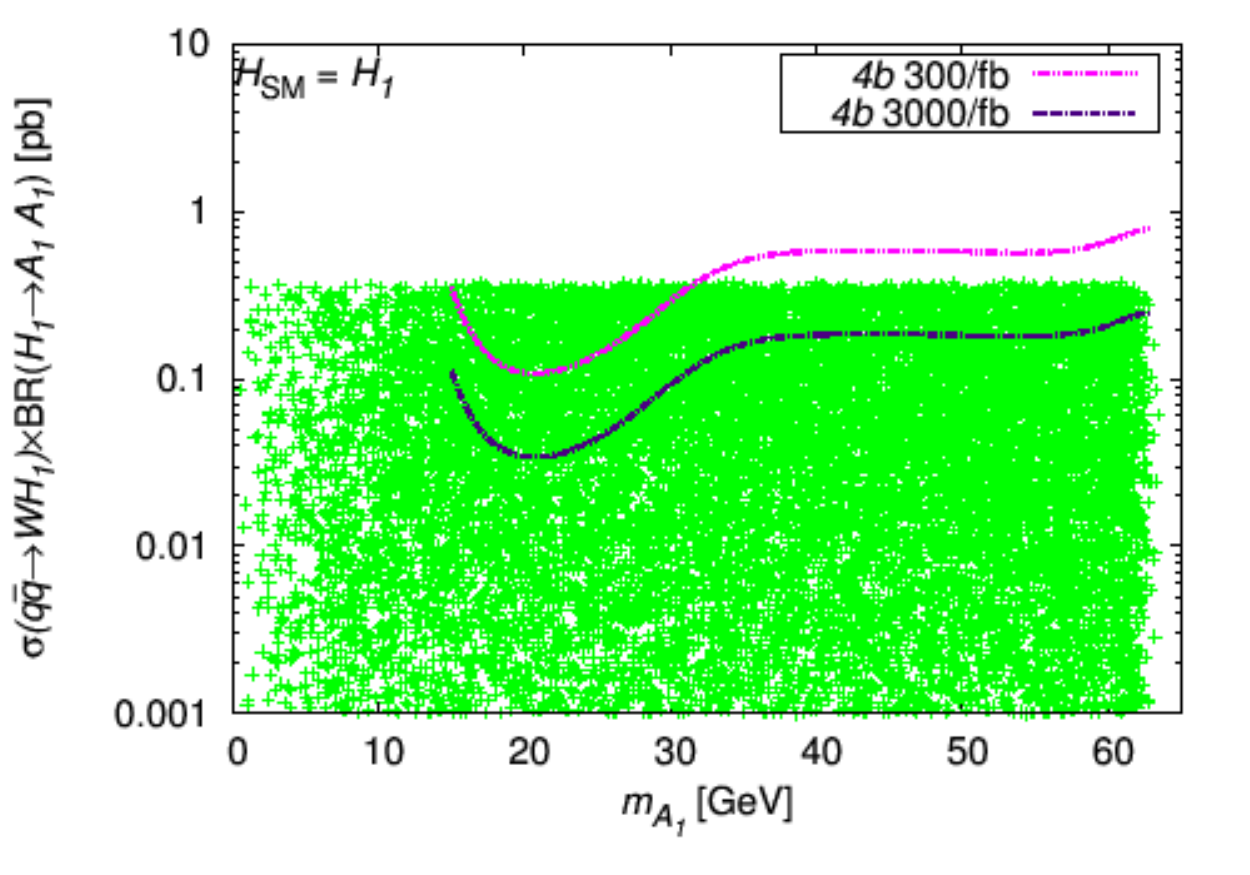}
}%
\hspace{0.5cm}%
\subfloat[]{%
\label{fig:-b}%
\includegraphics*[width=0.45\textwidth]{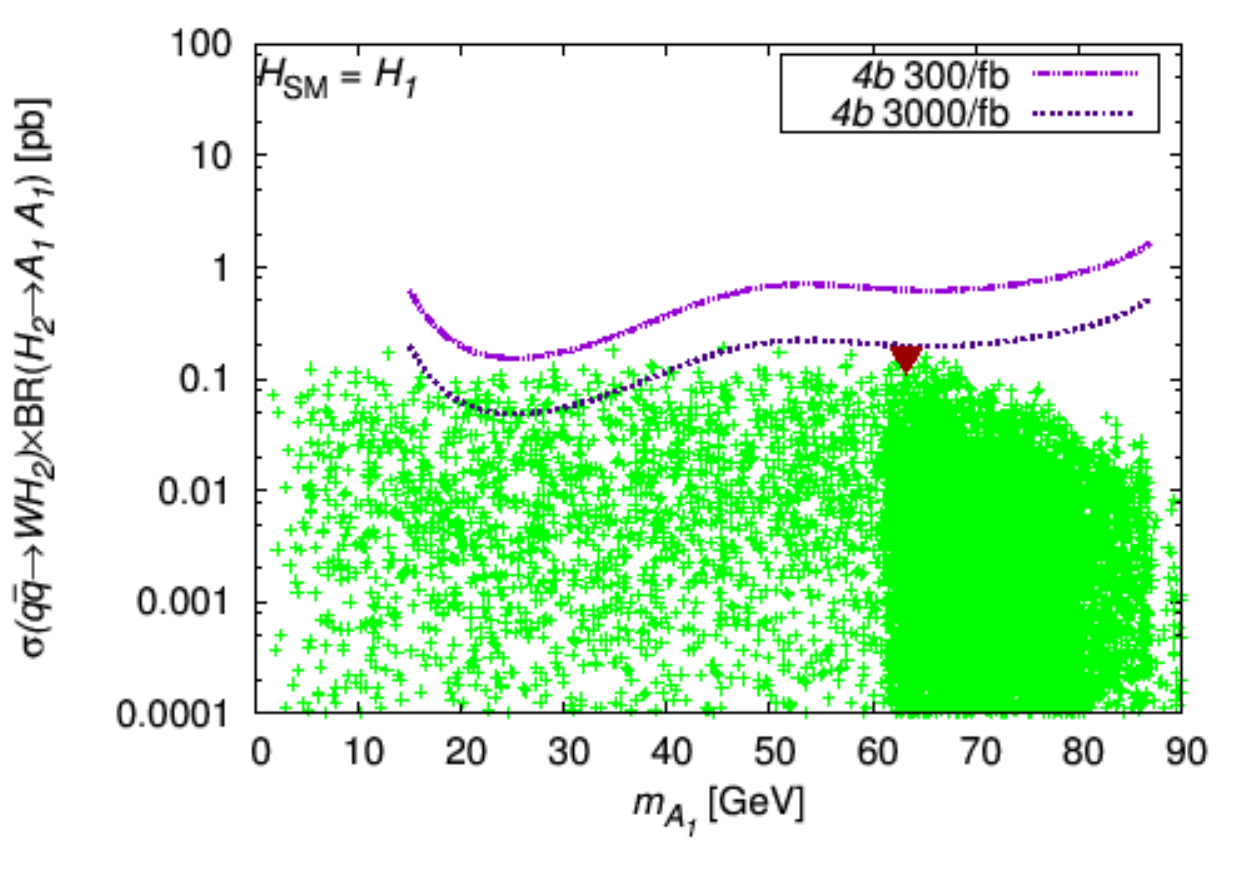}
}
\caption{LHC reach in $H_1\to A_1A_1$ (left) and $H_2\to A_1A_1$
  (right) for $H_1=H_{\rm SM}$ in the $WH$ channel. In panel (a) the sensitivity curves use \mhone\ =125 GeV, while (b) uses  $m_{H_2}=175$ GeV and both uses the best of the 4 $b$-jets and fat jets analyses. The colour code
  for the points is the same as in figure~\ref{fig:GF}.}
\label{fig:h1WH}
\end{figure}

In figure~\ref{fig:h1WH}(a) we show the discovery reach in the $H_1\to
A_1A_1$ channel in the $H_1=H_{\rm SM}$ case. The sensitivity curves
are set for $m_{H_1}=125$ GeV and the corresponding cut on the overall
invariant mass. If we compare with figure~\ref{fig:h1ZH}(a), the reach
here is much greater, even 300/fb might be enough for detecting some
$A_1$s around 20 GeV. Such a discovery is not possible even for VBF
production, as can be seen in figure~\ref{fig:h1VBF}(a), where 300/fb
does not reach the upper 1 pb limit for the rate. The reason for the relative
success of $WH$ production for these low masses is that in the $4b$
final state the use of fat jets leads to a more significant
improvement over a single $b$-jet analysis than is the case for the
$2b2\tau$ channel used for VBF production. Also, the backgrounds in the $WH$ channels are more severe at higher masses, rendering the low mass region more interesting.

The $H_2\to A_1A_1$ channel is much less optimistic, as can be seen in
figure~\ref{fig:h1WH}(b), where we set $\mhtwo =175$ GeV with no cut
on overall invariant mass for the sensitivity curves. It is clear the growth of the  $t\bar tb\bar b$ and $t\bar t$ backgrounds 
with increasing invariant masses suppresses the sensitivity at higher masses. One could
try some invariant mass cuts to suppress these backgrounds and possibly reach the points just above the $\maone=\mhone/2$ threshold,  but we
leave such considerations to section~\ref{benchmarks}.

\subsection{$\boldsymbol{H_2}$ SM-like}
\begin{figure}[tbp]
\centering
\subfloat[]{%
\label{fig:-a}%
\includegraphics*[width=0.45\textwidth]{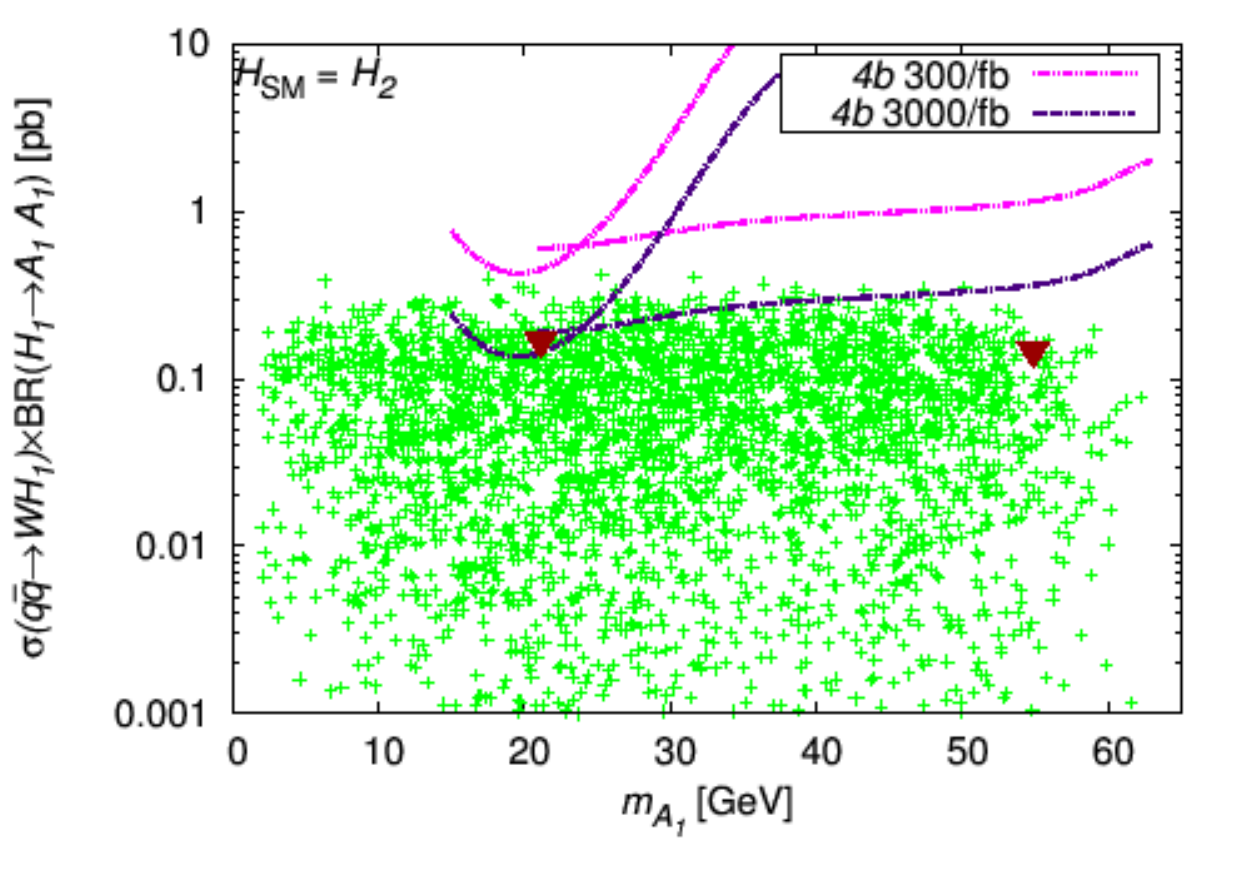}
}%
\hspace{0.5cm}%
\subfloat[]{%
\label{fig:-b}%
\includegraphics*[width=0.45\textwidth]{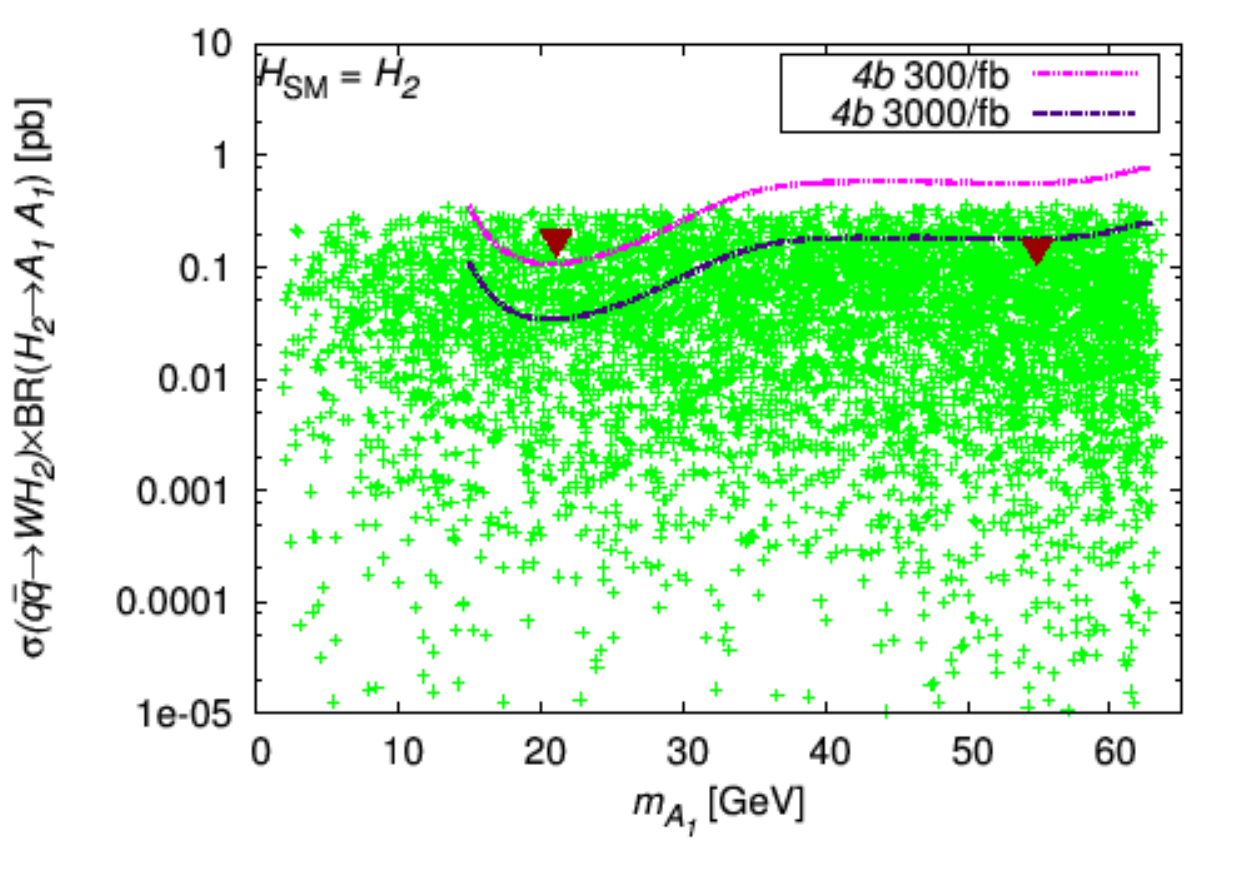}
}
\caption{LHC reach in $H_1\to A_1A_1$ (left) and $H_2\to A_1A_1$
  (right) for $H_2=H_{\rm SM}$ in the $WH$ channel. The sensitivity curves in panel (a) correspond to the two fat jets (the curves at low \maone\ using \mhone\ = 100 GeV) and the four single b-jets (the curves at higher \maone\ using \mhone\ = 125 GeV) analyses. In panel (b) \mhtwo\ = 125 GeV is used along with the best of the 4 $b$-jets and fat jets analyses. The colour code  for the points is the same as in figure~\ref{fig:GF}.}
\label{fig:h2WH}
\end{figure}

The case $H_2=H_{\rm SM}$ is rather similar to the $H_1=H_{\rm SM}$ case. As can be seen in figure~\ref{fig:h2WH}(a) ---
displaying sensitivity curves with $m_{H_1}=100$ GeV (using fat jets)
and $125$ GeV (using $b$-jets) but with no overall cut on invariant
mass --- there is just a marginal hope for detection in $H_1\to A_1A_1$ at 3000/fb. We do see improved sensitivity in the low
mass region as compared to VBF production, but not enough to cover any significant part of the parameter space. This is again a consequence
of the strength of the fat jet analysis for $4b$ as well as the main
backgrounds $t\bar tb\bar b$ and $t\bar t$ being smaller for lower
invariant masses.

For $H_2\to A_1A_1$ (figure~\ref{fig:h2WH}(b)) we again see essentially the same as figure~\ref{fig:h1WH}(a). Here we use $m_{H_2}= 125$ GeV and constrain
the four-body invariant mass of the final state to be $125\pm30$
GeV. The sensitivity curves use a combination of the fat jet and the
single $b$-jet analyses, always showing the more sensitive one.

\section{\label{benchmarks} Benchmark points}
As mentioned before, the detection sensitivities in all the channels
studied in this paper are significantly worse than the corresponding
results from GF production.\footnote{This is true unless the
  triggering in the GF channel turns out to be more challenging than
  what is presently hoped for.} Furthermore, this is generally true for
all parameter space regions that we have been able to access in the
analyses carried out here. Therefore it is fair to assume that, by the
time the channels under investigation become interesting for
experimental study, the lightest scalar ($H_1$ and $H_2$) as well as
pseudoscalar ($A_1$) Higgs states will already have been discovered
via GF and the goal for both the VBF and HS channels will become to
enable one further study their properties, as explained in the
introduction.

For this purpose we define three benchmark points and study them under
the assumption that we know the masses involved and can hence use this
information to further constrain the kinematics in the attempt to
increase the sensitivity.\footnote{Even without prior knowledge of
  the masses, these analyses can be employed by scanning over the
  masses involved. This, however, would render large look-elsewhere
  effects.} The details of the three points are given in
table~\ref{tab:Bench}. The points are chosen to cover as much as
possible of the parameter space within reach. Point 1 has a rather
light $A_1$ of 21 GeV and hence lies in the region where jet
substructure methods are of importance, while point 2 is closer to the
threshold for $H_{1,2}\to A_1A_1$ with $m_{A_1}=55$ GeV. The last
point is the only one with $H_1=H_{\rm SM}$ and with $m_{A_1}=63$ GeV,
it is designed for $H_2\to A_1A_1$ studies. Below the $\hsm\to A_1A_1$ threshold,
the phenomenology of \honesm\ and \htwosm\ is very similar, obviating any need for additional benchmark points in that region.

\begin{table}[tbp]
\begin{center}
\begin{tabular}{|c|c|c|c|}
\hline
Case & \multicolumn{2}{c|}{$H_{\rm SM} = H_2$} & $H_{\rm SM} = H_1$ \\
 \hline
& Point 1 & Point 2 & Point 3 \\
\hline
Input parameters & \multicolumn{3}{c|}{}\\
\hline
\mzero\,(GeV) & 1516.0 & 1470.5 & 1326.7 \\
\mhalf\,(GeV)  & 491.58 & 947.19 & 369.23 \\
\azero\,(GeV)  & $-2966.1$ & $2523.0$ & $2839.5$ \\
\mueff\,(GeV)  & 113.54 & 122.98 & 189.28  \\
\tanb &  13.051 & 12.46  & 2.882  \\
\lam & 0.08899 & 0.1641  & 0.5408  \\
\kap & 0.04042 & 0.07769 & 0.1818  \\
\alam\,(GeV) &  781.94 & $1714.9$ & $479.29$  \\
$M_p$,(GeV) & 21.77 & 59.38 & $68.52$ \\
\hline
Observables & \multicolumn{3}{c|}{}\\
\hline
$m_{A_1}$\,(GeV) & 21.12 & 54.87 & 63.26  \\
$m_{H_1}$\,(GeV) & 100.54 & 110.77 & 125.45  \\
$m_{H_2}$\,(GeV) & 125.22 & 125.31 & 138.90  \\
\hline
\end{tabular}
\caption{Some specifics of the three benchmark points.}
\label{tab:Bench}
\end{center}
\end{table}

\begin{table}[tbp]
\begin{center}
\resizebox{\textwidth}{!}{
\begin{tabular}{|c|c|c|c|c|c|c|c|}
\hline
  &  \multicolumn{4}{c|}{Point 1} & \multicolumn{2}{c|}{Point 2} &  Point 3\\\hline
  & \multicolumn{2}{c|}{$H_i=H_1$} & \multicolumn{2}{c|}{$H_i=H_2$} & $H_i=H_1$  & $H_i=H_2$  &  $H_i=H_2$ \\\hline
  & $b$-jets & fat jet & $b$-jets & fat jet & $b$-jets & $b$-jets & $b$-jets \\
\hline
$q\bar qH_i\to q\bar q2b2\tau$ & \multicolumn{7}{c|}{}\\\hline
Signal [pb] & $1.7\times 10^{-6}$ & $2.7\times 10^{-6}$ & $4.1\times 10^{-6}$ & $8.8\times 10^{-6}$ & $3.9\times 10^{-6}$ & $7.0\times 10^{-6}$ & $1.0\times 10^{-5}$\\
$2j+2b2\tau$ & $2.4\times 10^{-7}$ & $8.5\times 10^{-7}$ & $2.4\times 10^{-7}$ & $8.5\times 10^{-7}$ & $2.6\times 10^{-6}$ & $2.6\times 10^{-6}$ & $3.4\times 10^{-6}$\\
$2j+t\bar t$ & $6.8\times 10^{-7}$ & $2.2\times 10^{-6}$ & $6.8\times 10^{-7}$ & $2.2\times 10^{-7}$ & $1.1\times 10^{-6}$ & $1.1\times 10^{-6}$ & $1.5\times 10^{-6}$\\
$\mathcal{L}$ [fb$^{-1}$] & 8400 & 10000 & 2400 & 1100 & 5800 & 1900 & 1200 \\
\hline
$ZH_i\to 2\ell 4b$ & \multicolumn{7}{c|}{}\\\hline
Signal [pb] & $1.4\times 10^{-6}$  & $4.1\times 10^{-6}$ & $2.9\times 10^{-6}$ & $7.8\times 10^{-6}$ & $2.2\times 10^{-6}$ & $2.9\times 10^{-6}$ & $4.0\times 10^{-6}$\\
$Z+4b$ [pb] & $9.1\times 10^{-7}$ & $1.1\times 10^{-6}$ & $8.7\times 10^{-7}$ & $1.1\times 10^{-6}$ & $2.7\times 10^{-6}$ & $3.9\times 10^{-6}$ & $4.0\times 10^{-6}$\\
$\mathcal{L}$ [fb$^{-1}$] & 11000 & 2400 & 3500 & 1300 & 13000 & 11000 & 6300 \\
\hline
$WH_i\to \ell 4b$ & \multicolumn{7}{c|}{}\\\hline
Signal [pb] & $1.1\times 10^{-5}$ & $3.9\times10^{-5}$ & $2.1\times 10^{-5}$ & $7.3\times 10^{-5}$ & $1.5\times 10^{-5}$ & $2.0\times 10^{-5}$ & $2.7\times 10^{-5}$\\
$W+4b$ [pb] & $1.1\times 10^{-6}$ & $1.7\times 10^{-6}$ & $9.3\times 10^{-7}$ & $1.7\times 10^{-6}$ & $2.2\times 10^{-6}$ & $2.8\times 10^{-6}$ & $2.5\times 10^{-6}$\\
$t\bar tb\bar b$ [pb] & $8.2\times 10^{-6}$ & $1.4\times 10^{-5}$ & $7.5\times 10^{-6}$ & $1.8\times 10^{-5}$ & $3.1\times 10^{-5}$ & $4.5\times 10^{-5}$ & $4.2\times 10^{-5}$\\
$t\bar t$ [pb] & $9.3\times 10^{-7}$ & $4.9\times 10^{-7}$ & $1.3\times 10^{-6}$ & $6.9\times 10^{-7}$ & $4.3\times 10^{-6}$ & $7.7\times 10^{-6}$ & $1.1\times 10^{-5}$\\
$\mathcal{L}$ [fb$^{-1}$] & 2300 & 266 & 548 & 138 & 3900 & 3400  & 1900 \\
\hline
\end{tabular}}
\caption{Discovery prospects for  the three benchmark points. $\mathcal{L}$ denotes the integrated luminosity required for a detection in the given channel.}
\label{tab:Sigma}
\end{center}
\end{table}

In the study of these benchmark points we therefore use a somewhat modified kinematic analysis. Specifically, all $A_1$ candidates (i.e.,\ $b$-jet pairs, fat jets or $\tau$-jet pairs) are required to be within 15 GeV off the (assumed known) $A_1$ mass. For each point we run two simulations, one for $H_1\to A_1A_1$ and one for $H_2\to A_1 A_1$ and in each case the combined invariant mass of the two $A_1$ candidates is required to be within 30 GeV of the (assumed known) $H_1$ or $H_2$ mass.

The result of these studies are displayed in table~\ref{tab:Sigma},
where the cross sections after all cuts are presented for the signal
as well as the backgrounds. We also show the integrated luminosity
needed to obtain $S/\sqrt{B}>5$ with at least 10 events. The result of
the jet substructure (fat jet) analysis is only shown for point 1 as
this is the only scenario with an $A_1$ light enough for such studies
to be useful, though, in that case, this is clearly the most effective
approach. Also, we do not include $H_1\to A_1A_1$ for point 3 as this
channel is kinematically closed.

As stated before, the signals are in general larger than the
backgrounds and, in many cases, the main constraint on the required
luminosity is the requirement of at least 10 events. We also see in
table~\ref{tab:Sigma} that $WH$ is usually the most promising channel
as it has the highest rates. It also has the highest backgrounds
though, due to $t\bar tb\bar b$ and $t\bar t$ and, since these tend to
increase with increasing $H_i$ mass, we note that for $H_2\to A_1A_1$
in point 2 and 3 the VBF channel is somewhat better.

Comparing the detection prospects  in
table~\ref{tab:Sigma}, with the reach
shown in the corresponding plots, it is clear that the additional constraints on the
kinematics due to the assumption about the mass spectrum are 
important in improving sensitivity, though in some cases the differences are negligible. 
The invariant mass cuts are especially important for $WH$ as the
$t\bar tb\bar b$ and $t\bar t$ backgrounds increase significantly with
increasing overall invariant mass, hence rendering a (relatively low)
cut on over all invariant mass very effective.

\begin{figure}[tbp]
\centering
\subfloat[]{%
\label{fig:-a}%
\includegraphics*[width=0.45\textwidth]{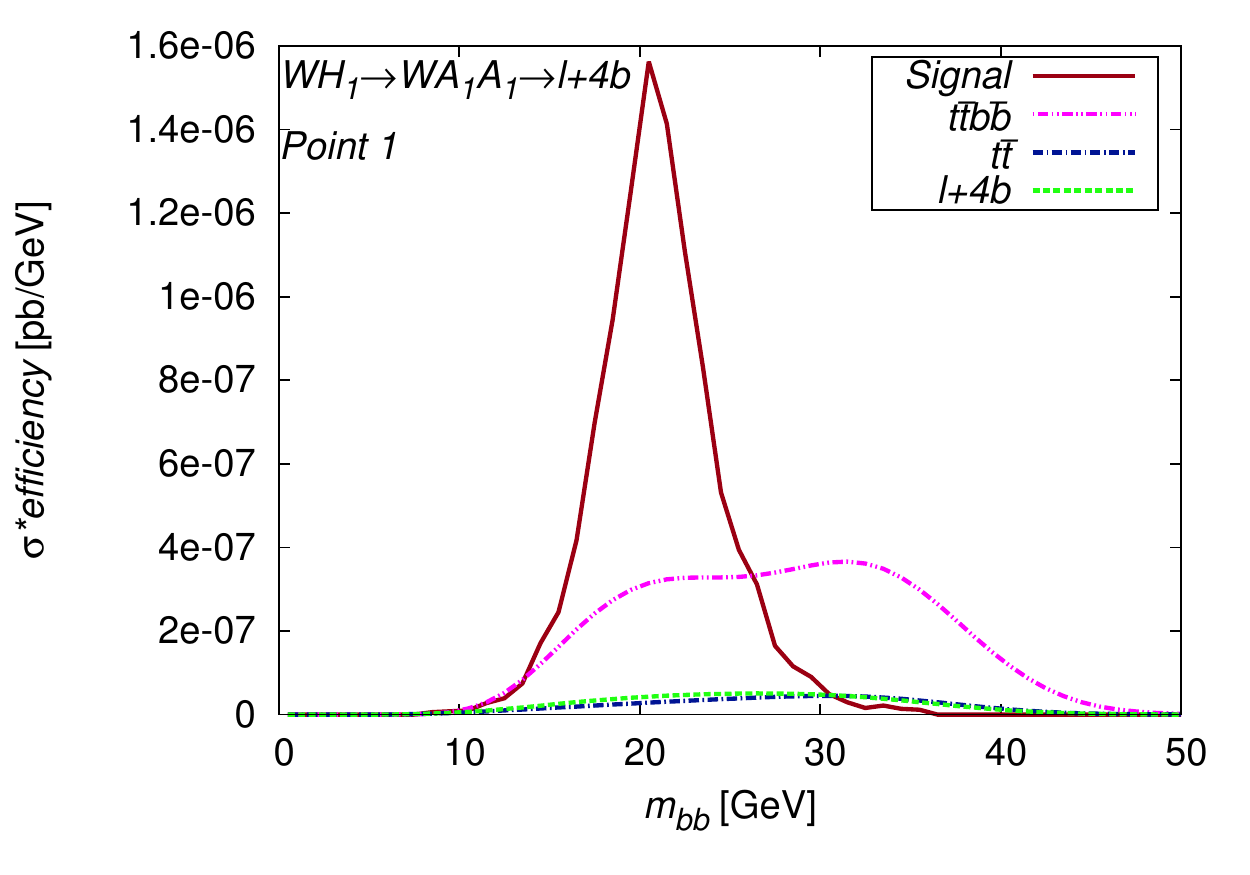}
}%
\hspace{0.5cm}%
\subfloat[]{%
\label{fig:-b}%
\includegraphics*[width=0.45\textwidth]{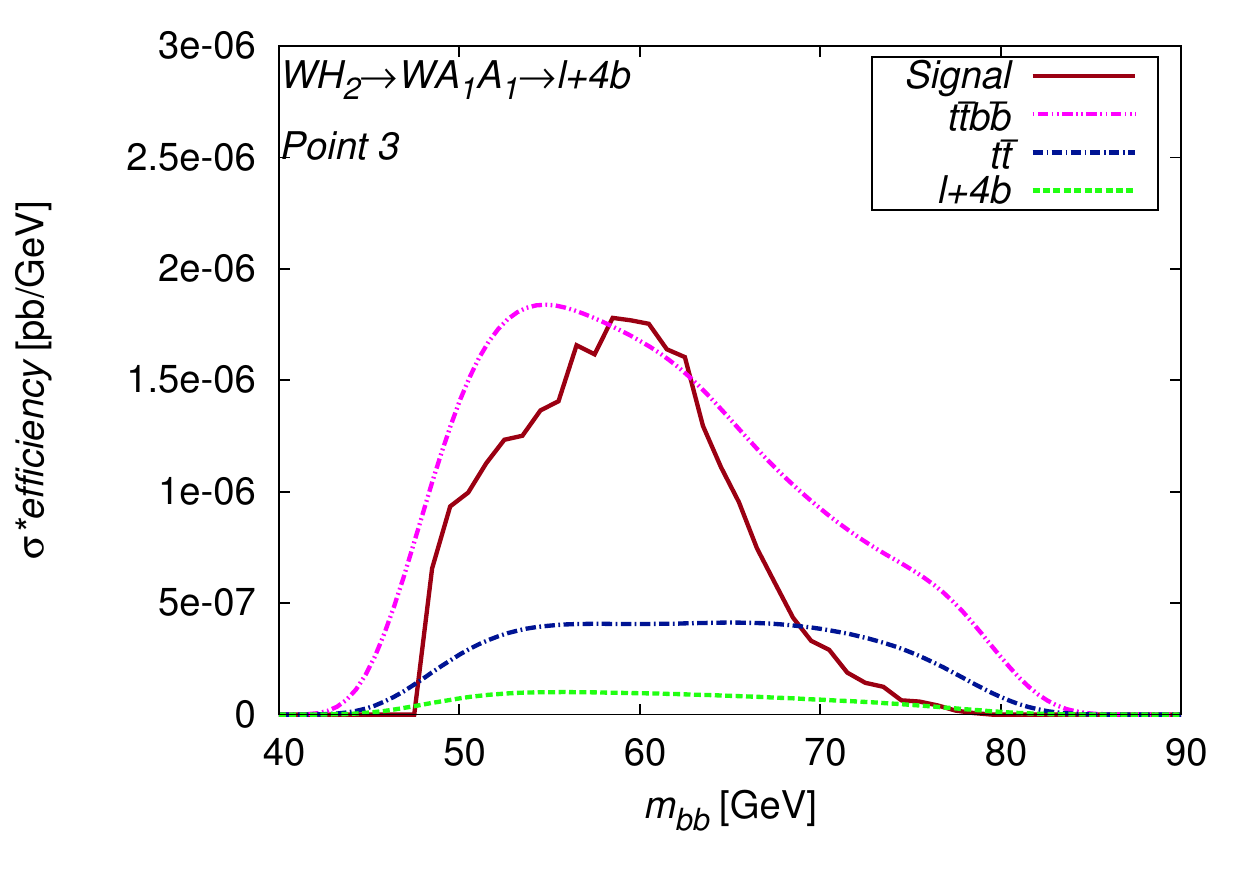}
}
\caption{Signal and backgrounds for two of our benchmark points in the $WH$ channel. The backgrounds have been smoothened for visual clarity.}
\label{fig:Bench}
\end{figure}

To illustrate more clearly the signal and backgrounds in the $WH$ channels
we plot these as functions of the invariant mass of the $b$-jet pairs\footnote{Included here are the two invariant masses of the combination of $b$-jet pairs that have the smallest difference in invariants mass of all the combinations of $b$-jet pairs where both pairs have  invariant mass within $\maone\pm 15$\ GeV.}, $m_{bb}$, in figure~\ref{fig:Bench} for
$H_1\to A_1A_1$ in point 1 (figure~\ref{fig:Bench}(a)) and for $H_2\to
A_1A_1$ in point 3 (figure~\ref{fig:Bench}(b)). Note that, 
due to the cuts in the analysis, the distributions are restricted to $m_{bb}\pm 15$ GeV and 
that both plots use single $b$-jet analysis only. In
figure~\ref{fig:Bench}(a) it is clear that a somewhat narrower cut
would significantly reduce the background from $t\bar tb\bar b$
without affecting the signal significantly. However, it is important
to note that there are significant statistical uncertainties in the
backgrounds (partially hidden by the smoothening) and this, together
with the fact that the signal already dominates and that we do not
include detector resolution, means that it is not clear how beneficial
such a cut would really be.
From figure~\ref{fig:Bench}(b) we see that for heavier masses the
invariant mass peak is smeared out towards smaller masses, this is mostly an effect of the difficulty in finding the ``right`' $b$-jet pairs; for lighter $A_1$s the ``wrong'' pairing gives too high invariant masses to be included.

\section{\label{epem} Prospects at Higgs factories}
Since we have seen that the conditions for detecting light pseudoscalars at the LHC via VBF and HS are rather challenging and it may require several years to do so, it is worth investigating whether a Higgs factory would do better in this respect. 
To this end, we estimate the sensitivity of a 240 GeV $e^+e^-$ collider that produces CP-even Higgses primarily 
through the $ZH$ channel. Since the hadronic background here is much smaller than at the LHC, we do not require any $b$-tagging but just look for a leptonically decaying $Z$-boson (into electrons/muons) accompanied by exactly 4 jets (each with $p_T>15$ GeV).

\begin{figure}[tbp]
\centering
\includegraphics*[width=0.45\textwidth]{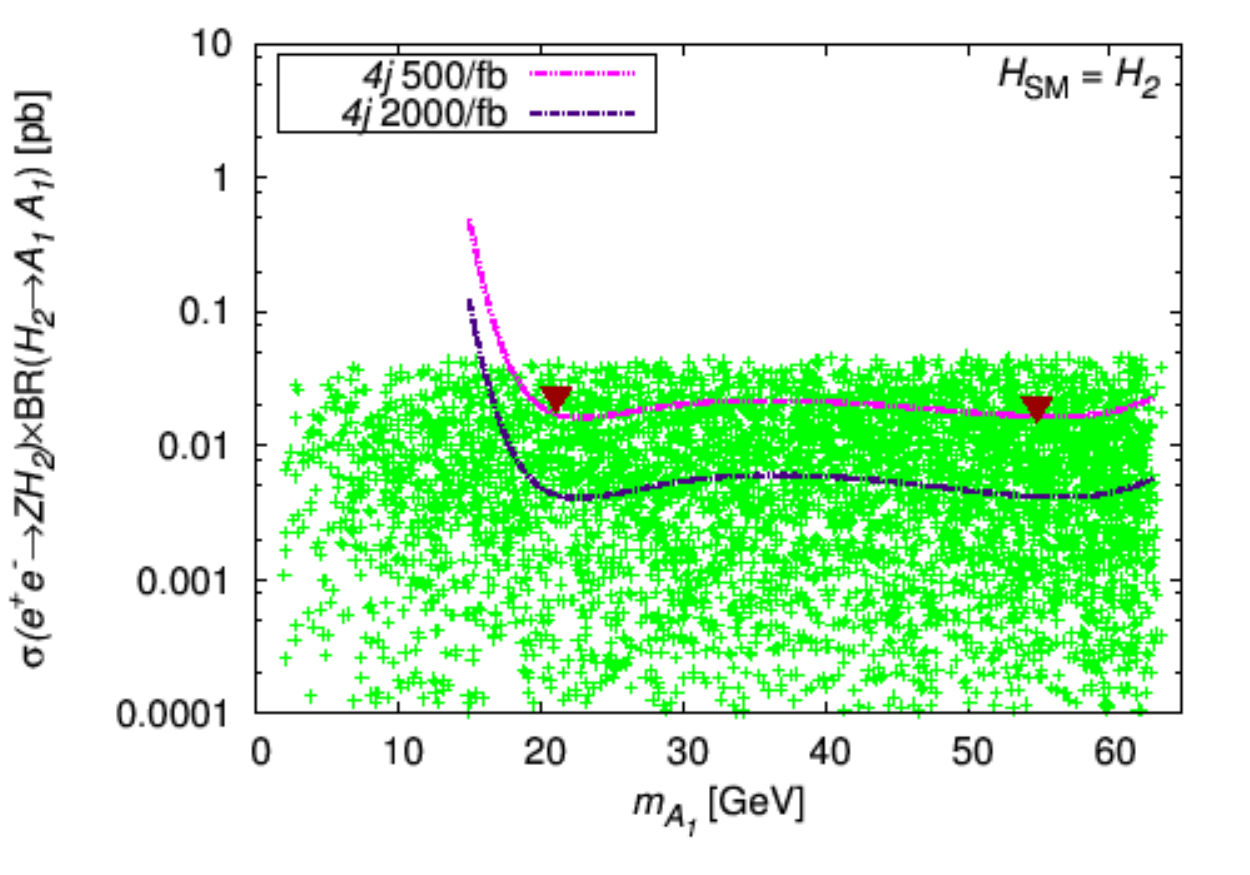}
\caption{Reach of a 240 GeV $e^+e^-$ collider in $H_2\to A_1A_1$
 for $H_2=H_{\rm SM}$ in the $ZH$ channel. All sensitivity curves assumes \mhtwo\ = 125 GeV and looks for four jets with total invariant mass of 125 GeV in addition to the leptonically decaying $Z$. The colour code
  for the points is the same as in figure~\ref{fig:GF}.}
\label{fig:TLEP}
\end{figure}

In all other aspects our analysis is identical to what was done for the LHC, except that we do not include jet substructure techniques at this point. In fact, we only study chains starting with \hsm\ for which we assume a production cross section of $\sigma(e^+e^-\to \hsm Z)=200$ fb. The resulting sensitivity is shown in figure~\ref{fig:TLEP}. It is clear that a Higgs factory would be much more suitable for our channels, as sensitivity to the NMSSM dynamics will be established already in the first year of operation. Furthermore, the very low background also means that profiling 
the  events  in terms of underlying mass spectra and {\hsm}$WW$ 
coupling strength would be an easier task than at the LHC.

\section{\label{conclusions} Conclusions}
In contrast to the more constrained MSSM, the NMSSM allows for the
existence of very light Higgs scalars as well as pseudoscalars. Therefore,
the discovery of, in particular, a light pseudoscalar Higgs state
would not just prove the existence of physics beyond the SM but would
also be inconsistent with minimal supersymmetry.

Since the NMSSM also accommodates the 125 GeV Higgs more naturally than the
MSSM, it is well worth a study. Although the most promising LHC discovery
channels of the aforementioned light pseudoscalar state are most
likely based on GF production of heavier scalars that subsequently
decay to $A_1A_1$ or $A_1Z$, we demonstrated here that also VBF and HS
production of the heavier scalars can be accessible. Hence, these two
additional production modes can be exploited to study couplings
not accessible in GF, such as those of the heavy scalars to SM gauge
bosons. Especially interesting are the channels starting with $H_S$:  as these can have BR$(H_i\to A_1A_1)$ close to 1, the channels of this paper might be our only chance of measuring the couplings of these scalars to gauge bosons.

In these channels, the signal rates are substantially lower than in
the case of GF, but the same is true for the backgrounds. Due to the
nature of the couplings involved, the only decay chains of interest
here are $H_{1,2}\to A_1A_1$. For VBF production of $H_{1,2}$ the most
promising of our final states is $2b2\tau$ (in addition to the two
forward/backward jets), which may allow detection, although only at
3000/fb.

For HS production, the background is even smaller, especially for the
$ZH$ mode. However, this channel also has a very small cross section
and hence the signal will be very hard to extract. As the event rates are
indeed significantly smaller compared to VBF and GF, here it is most
beneficial to employ the final state with the highest rate, i.e.,
$4b$.

Although still featuring a relatively low signal rate, $WH$ production
shows significantly better prospects than $ZH$. Despite significant
backgrounds from $t\bar tb\bar b$ and $t\bar t$, the higher signal
rates make this the most promising channel studied in this paper, at
least for relatively light initial scalars. However, given the
invariant mass structure of the main backgrounds, the signal tends to
be overwhelmed unless one can cut on the four-$b$ invariant mass and
the enforced mass window needs to be relatively narrow. Therefore, the
prospects for this channel diminish as the mass of the initially
produced scalar increases, rendering VBF the most promising channel
for heavier scalars.

In addition to general scans for sensitivity reach in parameter space,
we have performed more detailed studies of three representative
benchmark points. Here, we assumed knowledge of the masses of the
produced scalar as well as the pseudoscalar (e.g., as measured in GF
production) and used this information to constrain the
kinematics. Especially for HS, this can dramatically improve the
sensitivity.

Finally, we have briefly tested how the LHC prospects compare to those of a future
$e^+e^-$ machine running at the threshold of $ZH$ production, the golden (Bjorken in this case) 
channel of such a  Higgs factory. Needless to say, sensitivity to \hsm$\to A_1A_1$ decays is rather prompt herein.
Further, the clean environment of an $e^+e^-$ machine in terms of very limited backgrounds will enable precise
diagnostic of the NMSSM dynamics involved. It remain to be seen whether commissioning of such a leptonic machine
will occur on a timescale that will make it competitive with a potential LHC stage at high (instantaneous) luminosity,
the so-called Super-LHC (SLHC) option \cite{Gianotti:2002xx}, currently being discussed as a probable upgrade of the CERN collider past
Run II.

\begin{center}
  \textbf{Acknowledgments}
\end{center}

\noindent We would like to thank Shoaib Munir and Robin Aggleton for useful discussions.
This work has been funded in part by the Welcome Programme of the
Foundation for Polish Science. S.~Moretti is supported in part through
the NExT Institute. L.~Roszkowski is also supported in part by a
Lancaster-Manchester-Sheffield Consortium for Fundamental Physics
under STFC grant ST/L000520/1. The use of the CIS computer cluster at
NCBJ is gratefully acknowledged.

\bibliographystyle{h-physrev}	
\bibliography{NMSSM_A1_refs}

\end{document}